\documentclass[fleqn,usenatbib]{mnras}

\usepackage{newtxtext,newtxmath}
\usepackage[T1]{fontenc}

\DeclareRobustCommand{\VAN}[3]{#2}
\let\VANthebibliography\thebibliography
\def\thebibliography{\DeclareRobustCommand{\VAN}[3]{##3}\VANthebibliography}

\usepackage{graphicx}	% Including figure files
\usepackage{amsmath}	% Advanced maths commands
\usepackage{xcolor}
\usepackage{siunitx} %allow angles like arcseconds
\usepackage{booktabs}
\usepackage{spverbatim} %for codes text
\usepackage{verbatimbox}
\usepackage{journal_names} %Use journal names
\usepackage{comment}
\usepackage[flushleft]{threeparttable} %notes under table
\usepackage{caption}
\usepackage{subfigure}%for subfigures
\usepackage{longtable,booktabs}
\usepackage[online]{threeparttablex}
\usepackage{geometry,rotating}
\usepackage{url}
\usepackage{enumitem}
\setlist[itemize]{leftmargin=0.1in,topsep=0pt}
\usepackage{soul} %strikethrough text
\usepackage{pdflscape} %landscape
\usepackage{afterpage} %Should not cut text in landscape
\usepackage{placeins}
\usepackage{wrapfig}
\newcommand\hi{$\textrm{H}\,\scriptstyle\mathrm{I}$} %hi
\newcommand\hii{$\textrm{H}\,\scriptstyle\mathrm{II}$} %hii
\newcommand\HI{$\textbf{H}\,\scriptstyle\mathbf{I}$}
\newcommand\kms{$\rm km\,s^{-1}$}
\newcommand\mJy{$\rm mJy\,beam^{-1}$}
\newcommand{\msun}{M$_\odot$}
\defcitealias{Kraan-Korteweg2017}{KK2017}

\title[Unveiling the hidden VSCL with Vela$-$SMGPS]{\HI\ Galaxy Signatures in the SARAO MeerKAT Galactic Plane Survey $-$ III. Unveiling the obscured part of the Vela Supercluster}

\author[S. H. A. Rajohnson et al.]{
Sambatriniaina H. A. Rajohnson$^{1}$,\thanks{E-mail: aychasam@gmail.com}
Renée C. Kraan-Korteweg$^{1}$,
Hao Chen$^{1,2}$,
Bradley S. Frank$^{3,4,5,1}$,
\newauthor{Nadia Steyn$^{1,6}$,
Sushma Kurapati$^{1}$,
D. J. Pisano$^{1}$,
Lister Staveley-Smith$^{6}$,
Paolo Serra$^{7}$, Sharmila}
\newauthor{Goedhart$^{3,8}$,
Fernando Camilo$^{3}$}
\\
$^{1}$Department of Astronomy, University of Cape Town, Private Bag X3, Rondebosch 7701, South Africa\\
$^{2}$Research Center for Intelligent Computing Platforms, Zhejiang Laboratory, Hangzhou 311100, China\\
$^{3}$South African Radio Astronomy Observatory (SARAO), 2 Fir Street, Observatory, 7925, South Africa\\
$^{4}$UK Astronomy Technology Centre, Royal Observatory Edinburgh, Blackford Hill, Edinburgh EH9 3HJ, UK\\
$^{5}$The Inter-University Institute for Data Intensive Astronomy (IDIA), and University of Cape Town, Private Bag X3, Rondebosch, 7701, South Africa\\
$^{6}$International Centre for Radio Astronomy Research (ICRAR), University of Western Australia, Crawley, WA 6009, Australia\\
$^{7}$INAF – Osservatorio Astronomico di Cagliari, Via della Scienza 5, 09047, Selargius (CA), Italy\\
$^{8}$SKAO, 2 Fir Street, Black River Park, Second Floor, Block A, Cape Town, 7925\\
}

\date{Accepted 2024 May 24. Received 2024 May 22; in original form 2024 February 2}

\pubyear{2022}

\begin{document}
\label{firstpage}
\pagerange{\pageref{firstpage}--\pageref{lastpage}}
\maketitle

\begin{abstract}
We conducted a search for \hi{} emission of the gas-rich galaxies in the Vela region ($260^{\circ} \leq \ell \leq 290^{\circ}, -2^{\circ} \leq b \leq 1^{\circ}$) to explore the Vela Supercluster (VSCL) at $V_\mathrm{hel} \sim 18000$ \kms{}, largely obscured by Galactic dust. Within the mostly RFI-free band ($250 < V_\mathrm{hel} < 25000$ \kms{}) of MeerKAT, the analysis focuses on 157 hexagonally distributed %MeerKAT 
pointings extracted from the SARAO MeerKAT Galactic Plane Survey located in the Vela region (Vela$-$SMGPS). These were combined into 10 contiguous mosaics, covering a ${\sim}90$ deg$^2$ area. Among the 843 \hi{} detected sources, 39 were previously discovered in the Parkes HIZOA survey ($V_\mathrm{hel} < 12000$ \kms{}; rms $\sim$ 6 \mJy{}). With the improved rms level of the Vela$-$SMGPS, i.e., $0.29 - 0.56$ \mJy{}, our study unveils nearly 12 times more detections (471 candidates) in that same velocity range. We furthermore could identify 187 galaxy candidates with an \hi{} mass limit reaching $\log (M_{\rm HI}/\rm M_{\odot}) = 9.44$ in the VSCL velocity range $V_\mathrm{hel} \sim 19500 \pm 3500$ \kms{}. We find indications of two wall-like overdensities that confirm the original suspicion that these walls intersect at low latitudes around longitudes of $\ell \sim 272^{\circ} - 278^{\circ}$. We also find a strong signature most likely associated with the Hydra/Antlia extension and evidence of a previously unknown narrow filament at $V_\mathrm{hel} \sim 12000$ \kms{}. This paper demonstrates the efficiency of systematic \hi{} surveys with the SKA precursor MeerKAT, even in the most obscured part of the Zone of Avoidance (ZOA).
\end{abstract}


%%%%%%%%%%%%%%%%%%%%%%%%%%%%%%%%%%%%%%%%%%%%%%%%%%

%%%%%%%%%%%%%%%%% BODY OF PAPER %%%%%%%%%%%%%%%%%%
\section{Introduction}\label{sec:intro}
On large scales, galaxies are arranged in a web-like pattern comprising groups, filamentary structures with high-density clusters at their nodes, surrounding low-density voids -- known as the `cosmic web' \citep{Bond1996}. Due to the gravitational instability exerted by these mass variations on their surrounding medium, these overdense -- and underdense -- regions induce a peculiar velocity or cosmic flow field that deviates from a pure Hubble expansion \citep[e.g.,][]{Peebles1976,Lauer1994,Strauss1995}. The Cosmic Microwave Background (CMB) dipole, for example, is explained by the peculiar motion of the Local Group (LG) with respect to the CMB with a velocity of around $620 \pm 15$ \kms{} and an apex at $(\ell, b) = 271.9^{\circ} \pm 2.0^{\circ}, 29.6^{\circ} \pm 1.4^{\circ}$ \citep{Planck2020}. 

Numerous efforts have been made to delineate large-scale structures (LSS) and study cosmic flow fields in the nearby Universe thanks to research using data from (i) all-sky systematic redshift (e.g., CfA/CfA2, \citealt{Huchra1983,Falco1999}; SSRS2, \citealt{dacosta1998}; 2dFGRS, \citealt{Colless2001,Colless2003}; 6dFGRS \citealt{Jones2009}; 2MRS, \citealt{Huchra2012,Macri2019}; SDSS, \citealt{Tempel2014,Ahumada2020}) and (ii) peculiar velocity surveys (e.g., SFI++, \citealt{Masters2006,Springob2007}; 2MTF, \citealt{Masters2007,Springob2016,Hong2019}; 6dFGSv, \citealt{Springob2014,Scrimgeour2016}; Cosmicflows-3 or CF3, \citealt{Tully2016}; SDSS PV, \citealt{Howlett2022}; and Cosmicflows-4 or CF4 \citealt{Kourkchi2020,Tully2023}).\\

In spite of these advancements, neither the local bulk flow direction nor the volume 
as determined independently from these large data sets are fully coherent with the CMB dipole measurements \citep[e.g.,][]{Erdogdu2006a,Erdogdu2006b,Lavaux2010,Hoffman2017,Howlett2022}. Discrepancies persist in the literature (see e.g., Table 5 in \citealt{Scrimgeour2016}). On scales less than $150 \, \rm h^{-1}$\,Mpc, some studies show unusually large bulk flow amplitudes with respect to $\Lambda$CDM predictions  \citep{Hudson2004,Erdogdu2006a,Erdogdu2006b,Watkins2009,Feldman2010,Lavaux2010,Howlett2022}, while others present lower or consistent values \citep{Colin2011,Branchini2012,Turnbull2012,Hoffman2015,Scrimgeour2016,Qin2018,Qin2019,Qin2021}. 

The so-called ``Zone of Avoidance'' (ZOA; \citealt{proctor1878,Shapley1961,Kraan_lahav2000}), is considered to be a major contributor to these controversies \citep{Loeb2008} due to the incomplete mapping of the sky at low Galactic latitudes. Deep optical surveys in the ZOA (mostly photometric, see \citealt{Kraan_lahav2000,Kraan2005} for reviews) have significantly reduced its extent by covering lower latitudes (out to $|b| \leq 5^{\circ})$ and higher levels of extinction ($A_B \leq 3.\kern-0.5ex^{\mathrm{m}}0$). These surveys unveiled nearby dynamically significant structures such as the Norma cluster \citep{Woudt1998} the Puppis cluster \citep{Lahav1993}, and the Ophiuchus cluster \citep{Hasegawa2000,Wakamatsu2005} out to $V_\mathrm{hel} < 9000$ \kms{}. Despite these efforts, the ZOA gap persists, affecting about 20\% of the sky in optical wavelengths. 

This degree of obstruction diminishes, but not entirely, as we move towards longer wavelengths, e.g., Infrared (IR) wavelengths are less affected by extinction. For instance, the 2MASS Redshift Survey (2MRS), provided systematic follow-up redshifts across the entire sky for all the brightest galaxies from the near-IR (NIR) 2MASS Extended Source Catalog (2MASS XSC; \citealt{Jarrett2000}). This survey NIR redshift is 97.4\% complete down to $K_s \leq 11.75$ mag and $|b| \geq 5^{\circ}$, i.e., leaving 10\% of the sky uncharted in a systematic way. Obtaining redshifts for these partly obscured red galaxies is very hard, if not impossible. It was only recently that \cite{Macri2019} have added 1041 new redshift measurements to the 2MRS NIR catalog to complete it, reducing the previous lack of redshift information for galaxies located near the ZOA.

A new avenue now exists with data from the Wide-field Infrared Survey Explorer catalog (WISE, \citealt{Wright2010,Cutri2014}) in the mid-IR band. The most recent efforts in addressing the ZOA gap have been undertaken by \cite{Daza-Perilla2023}, who introduced an automated classification system to distinguish galaxies from non-galaxies in low Galactic latitude regions using data from the VVV NIR Galaxy Catalogue \citep{Baravalle2021}.

In addition to the incomplete mapping of galaxies behind the ZOA, bulk flow measurements conducted so far face constraints due to the volumes they have surveyed. Bulk flows derived from the IRAS Point Source catalog redshift survey (PSCz; \citealt{Hudson2004}), the 6dF Galaxy Survey (6dFGS; \citealt{Springob2014}), and the 2MASS Tully-Fisher survey (2MTF; \citealt{Hong2014,Hong2019}) have only considered structures below 10000, 12000, and 16000 \kms{}, respectively. The more recent surveys hint at a `residual bulk flow' originating from a hidden mass overdensity behind the ZOA \citep[cf.][]{Hudson2004,Loeb2008, Boruah2020}. 
For example, the 6dFGS and 2MTF surveys implied a residual bulk flow of ${\sim}273$ \kms{} towards the constellation of Vela ($\ell \sim 270^{\circ}-330^{\circ}$) in the ZOA and induced beyond $V_\mathrm{hel} > 16000$ \kms{} \citep{Springob2014,Springob2016,Scrimgeour2016}. 

Based on multi-object spectroscopic observations, \cite{Kraan-Korteweg2017}, hereafter \citetalias{Kraan-Korteweg2017}, recently uncovered the Vela Supercluster (VSCL). VSCL, centered at $\ell = 272.5^{\circ} \pm 20^{\circ}$ and $ b=0^{\circ}\pm 10^{\circ}$, hence in close proximity to the volume range and the direction over which this residual bulk flow motion is expected to arise. Using data from the Southern African Large Telescope (SALT), followed by data from AAOmega+2dF, \citetalias{Kraan-Korteweg2017} have uncovered a massive concentration of galaxies at a redshift of $V_\mathrm{hel} \sim 18000$ \kms{}. This concentration extends over $115 \times 90 \, h_{70}$ Mpc at this distance, with an estimated overdensity of $\delta \sim 0.50-0.77$. The inspection of VSCL's distribution in redshift space revealed a multi-branching morphology consisting of two wall-like structures. A main wall that is visible on both sides of the Galactic Plane (GP) at 18000 \kms{}, and a second smaller one below the GP at 22000 \kms{}. They seem to cross in the inner ZOA, which is unfortunately obscured. Although VSCL's inner part could not be traced because of the high extinction and high stellar crowding in the optical and NIR at the lowest latitudes ($|b| \leq 5^{\circ}$), a contribution to the motion of the LG based on a complete set of 2MASS galaxies on either side of the GP between $6^{\circ} \leq |b| \leq 10^{\circ}$ is estimated to be V$_{\rm LG} \simeq 50$ \kms{} (\citetalias{Kraan-Korteweg2017}). VSCL could therefore play a crucial role in explaining the residual bulk flow and refining our understanding of the Universe's dynamics. Further kinematic evidence based on reconstructed density and velocity field data reinforces the presence of VSCL with the two merging walls hiding a dense core (see \citealt{Sorce2017,Courtois2019}).

As a result, it is essential to trace VSCL across the ZOA, including areas of high obscuration, to gain a better understanding of its extent and mass overdensity. The most effective method to overcome this challenge is systematic \hi{} surveys, given that the 21-cm \hi{} emission passes unhindered through the ZOA. Previous single-dish surveys such as the Dwingeloo Obscured Galaxy Survey (DOGS, \citealt{Rivers1999}), the \hi{} Parkes Zone of Avoidance survey (HIZOA, \citealt{Staveley2016}), the deep Arecibo L-band Feed Array Zone of Avoidance survey (ALFAZOA, \citealt{McIntyre2015}), or interferometric surveys such as the Westerbork Radio Synthesis Telescope survey of the Perseus-Pisces filament (WSRT PP, \citealt{Ramatsoku2016}) have demonstrated their capabilities in mapping structures bisected by the ZOA. Yet, none of these surveys had the required sensitivity, velocity coverage, and field of view to effectively map the VSCL. This makes MeerKAT ideal for observing the VSCL. A pilot project using MeerKAT16 data has already successfully mapped the galaxy cluster VC04 ($\ell \sim 272.25^{\circ},~b \sim -9^{\circ}$, $V_\mathrm{hel} \sim 18000$ \kms{}) embedded in the main VSCL wall \citep{Steyn2023_Msc}. With a mean rms noise of 1.2 \mJy{}, the 6 pointings have led to the detection of 119 galaxies. 

In this paper, we use \emph{interferometric} \hi{} data obtained as part of the SARAO MeerKAT Galactic Plane survey (referred to as SMGPS) to investigate the \hi{} emission from the gas-rich galaxy population within the densest layer of the Milky Way plane and focus on learning more about the crossing of the VSCL walls. Firstly, the better spatial resolution and sensitivity compared to the previously undertaken surveys in the ZOA, the L-band frequency coverage, and the field of view of SMGPS will enable us to extend the depth of our study to 25000 \kms{}. Secondly, the SMGPS footprint encompasses the VSCL region. This paper presents the first result of an extensive exploration with a particular emphasis on mapping the uncharted inner core of the VSCL and uncovering possible intersections between its two walls through systematic \hi{} surveys with MeerKAT. A dedicated survey, linking this finding to known structures beyond the ZOA will be detailed in an upcoming publication (Rajohnson et al. in prep.). By detecting a significant portion of the star-forming, gas-rich population within the VSCL, we aim to derive a qualitative estimation of the supercluster's mass overdensity and gain insights into its contribution to the local bulk flow.

This paper is structured as follows: in Section \ref{sec:reduction}, we describe the observation details, data reduction processes, and mosaicking strategies of Vela$-$SMGPS. We explain the methods used to detect sources and verify the measured fluxes in Section \ref{sec:sofia}. Section \ref{sec:cat} presents %the new catalog and 
an overview of the source finding results of the highly-obscured galaxies and includes a completeness estimation. 
This is followed by a comparison of the \hi{} properties with predictions based on simulations in Section \ref{sec:properties}, and a detailed discussion of the newly discovered LSS crossing the ZOA in Section \ref{sec:lss}. Finally, we summarize and conclude our findings in Section \ref{sec:conclusion}.

Throughout this paper, we use the heliocentric velocity in optical convention $V_{\mathrm{hel}}$, H$_0 = 70$ \kms{}\,Mpc$^{-1}$, $\Omega_{\rm m} = 0.3$ and $\Omega_{\Lambda} = 0.7$ \citep{Hinshaw2013}.

\section{MeerKAT Observations and Data processing}\label{sec:reduction}
\subsection{The SARAO MeerKAT Galactic Plane Survey}\label{subsec:smgps}

SMGPS is an L-band (856-1712 MHz) continuum survey centered at 1.3 GHz, covering the longitude range $251^{\circ} \leq \ell \leq 61^{\circ}$ within a narrow strip with a width of $\Delta b = 3^{\circ}$ centered along the GP. With approximately 900 pointings, the latitude range was slightly adjusted to approximate the warp in the Milky Way disc. Within the longitude range $298^{\circ} \leq \ell \leq 61^{\circ}$, the corresponding latitude range covered is $-1.5^{\circ} \leq b \leq 1.5^{\circ}$. In contrast, within the longitude range of $251^{\circ} \leq \ell \leq 298^{\circ}$, the survey encompasses the latitude range $-2^{\circ} \leq b \leq 1^{\circ}$.

The primary objective of SMGPS is to study the continuum emission of the Galactic populations, e.g., Supernovae remnants, Planetary Nebulae, \hii{} regions, etc. (for more details, see \citealt{Goedhart2023}). However, its technical capabilities also enable the exploration of the hidden extragalactic LSS located behind the innermost Milky Way, such as the Great Attractor ($302^{\circ}\leq \ell \leq 332^{\circ}$, GA$-$SMGPS, \citealt{Steyn2023_paper}), and the Local Void ($328^{\circ} \leq \ell \leq 55^{\circ}$, LV$-$SMGPS, \citealt{Kurapati2023}), using its \hi{} data. Our region of interest in this work is centered on the proposed crossing of the VSCL at $\ell \sim 275^{\circ}$, Vela$-$SMGPS. It spans $30^{\circ}$ in longitude, extending from $260^{\circ} \leq \ell \leq 290^{\circ}$, with $-2^{\circ} \leq b \leq 1^{\circ}$, covering a $30^{\circ} \times 3^{\circ}$ (90 deg$^2$) area.

\subsection{Observations and data reduction}\label{subsec:reduction}
The deep observations (${\sim}$1hr per pointing) were performed between July 2018 and March 2020, using the full MeerKAT array, in full polarization. 
Each observing block of Vela$-$SMGPS, i.e., the set of data collected during one observation session, comprises 9 MKT pointings arranged in a hexagonal pattern. The pointing centers were offset by $0.494^{\circ}$ to reach uniform sensitivity. Table \ref{tab:obs} provides details of the analyzed region. It contains 157 individual fields observed during 19 blocks of observations, with a minimum of 58 antennas up to a maximum of 63 per observation. Due to scheduling constraints, two of these blocks were identical but split over multiple nights of observations. The observations were obtained %using the L-band 
with the 4K correlator mode and 8-second integration \citep{Goedhart2023}. With approximately 1 hour of on-source time per pointing, a session could last from 8.54 to 12.25 hours of on-track time. Typically, the complex phase calibrator (secondary calibrator) was visited for 65 seconds every 30 minutes after a cycle between the targets (${\sim}3$ min per field), and the bandpass and flux calibrators (primary calibrators) were scanned for 5 minutes every 3 hours. \\

\begin{table*}
    \small
	\centering
	\caption{Overview of Vela$-$SMGPS observational parameters}
	\label{tab:obs}
		\begin{tabular}{@{} c c c r l c c @{}}
		    \hline
			\hline
			Date & Block ID & Block centre & Track & N$_{\rm ant}$ & Primary & Secondary calibrators \\
			(UT start) &    & ($\ell, b$) & (h)   &   & calibrator & \\
			\hline
		    2018-12-30 &	1546207810 & (291.8, \,\,0.0) &	12.02 &	62 &	1215-457    &	J1331+3030, J1939-6342 \\
		    2018-12-31 &	1546293818 & (291.3, -1.5) &	12.25 &	62 &	1215-457 &	J1331+3030, J1939-6342 \\
            2019-01-01 &	1546380746 & (288.4, -0.5)	& 	11.99 &	62 &	1215-457 &	J1331+3030, J1939-6342 \\
            2019-01-04 &	1546641291 & (285.4, \,\,0.5)	&	10.51 &	61 &	J0906-6829 &	J0408-6545, J1331+3030, J1939-6342 \\
            2019-01-05 &	1546727810 & (285.9, -1.0)    &	10.39 &	60 &	J0906-6829 &	J0408-6545, J1331+3030, J1939-6342 \\
            2019-01-06 &	1546811642 & (282.9, \,\,0.0)	&	10.00 &	60 &	J0906-6829 &	J0408-6545, J1331+3030, J1939-6342 \\	
            2019-01-07 &	1546900349 & (282.4, -1.5)	&	8.54 &	62 &	J0906-6829 &	J0408-6545, J1331+3030, J1939-6342 \\
            2019-01-08 &	1546984998 & (279.5, -0.5)	&	9.97 &	62 &	1215-457 &	J1331+3030, J1939-6342 \\
            2019-01-09 &	1547071945 & (276.5, \,\,0.5)	&	9.32 &	62 &	1215-457 &	J1331+3030, J1939-6342 \\
            2019-01-12 &	1547330621 & (277.0, -1.0)	&	9.31 &	62 &	J0825-5010 &	J0408-6545, J1331+3030, J1939-6342 \\
            2019-01-13 &	1547413577 & (274.0, \,\,0.0)	& 	9.31 &	62 &	J0825-5010 &	J0408-6545, J1331+3030, J1939-6342 \\
            2019-01-14 &	1547499081 & (273.6, -1.5)	&	9.72 &	62 &	J0825-5010 &	J0408-6545, J1331+3030, J1939-6342 \\
            2019-01-15 &	1547585175 & (270.6, -0.5)	&	9.75 &	63 &	J0825-5010 &	J0408-6545, J1331+3030, J1939-6342 \\
            2019-01-17 &	1547758692 & (267.6, \,\,0.5)	&	9.31 &	58 &	J0825-5010 &	J0408-6545, J1331+3030, J1939-6342 \\
            2019-01-18 &	1547839876 & (268.1, -1.0)	&	10.01 &	63 &	J0825-5010 &	J0408-6545, J1331+3030, J1939-6342 \\
            2019-11-16 &	1573939357 & (265.2, \,\,0.0)	&	12.03 &	61 &	J0825-5010 &	J0408-6545, J0521+1638, J1331+3030, J1939-6342 \\
            2019-11-15 &	1573852860 & (264.7, -1.5)	&	12.02 &	61 &	J0825-5010 &	J0408-6545, J0521+1638, J1331+3030, J1939-6342 \\
            2019-11-22 &	1574456164 & (261.7, -0.5)	&	11.67 &	60 & 	J0825-5010 &	J0408-6545, J0521+1638, J1331+3030, J1939-6342 \\
            2019-11-23 &	1574542866 & (259.7, \,\,0.5)	& 	10.89 &	60 &	J0825-5010 &	J0408-6545, J0521+1638, J1331+3030 \\
            2019-11-26 &	1574801265 & (259.2, -1.0)	&	9.63 &  63 &	J0825-5010 &	J0408-6545, J0521+1638, J1331+3030 \\
            2019-11-28 &	1574911560 & (259.2, -1.0)	& 	2.39 &	61 &	J0825-5010 &	J0408-6545, J1331+3030 \\
			\hline
		\end{tabular}
	\end{table*}

We transferred each relevant block (see Table \ref{tab:obs}) from the SARAO archive\footnote{https://archive.sarao.ac.za/} into Ilifu\footnote{www.ilifu.ac.za}, consisting of approximately 211 hours of total observing time. Throughout this process, autocorrelation data were excluded, and flags generated by the monitoring system of the Telescope were automatically applied. Using orthogonal polarization components (HH/VV) only, we processed the frequency range 1308 to 1430 MHz ($ -2000  \leq V_\mathrm{hel} \leq 26 000$ \kms{}), which is mostly free from Radio Frequency Interference (RFI), and omitted the frequency range associated with Milky Way \hi\ emission, spanning from 1419.4 to 1421.4 MHz. This approach allows us to explore our volume of interest which is just at the border above which RFI becomes prominent. It covers the LSS out to $V_\mathrm{hel} < 25 000$ \kms{}, encompassing the entire redshift range for the study of gas-rich spirals at the supercluster distance (approximately $V_\mathrm{hel} \sim 16 000$ \kms{} to about 23000 \kms{}), and constitutes a scientifically relevant volume for bulk flow studies (e.g., \citealt{Scrimgeour2016}).\\

We used the CARACal pipeline \citep{Jozsa2020} which is specifically designed for continuum and spectral line radio interferometry data reduction. CARACal operates a collection of \textsc{Stimela}\footnote{https://github.com/SpheMakh/Stimela} scripts \citep{Makhathini2018} -- a framework based on Python and container technologies -- which allows various radio interferometry software to make the calibration and imaging workflow running sequentially. RFI flagging was performed through \texttt{Tricolour}\footnote{https://github.com/ratt-ru/tricolour} for the primary calibrator flagging and \texttt{Tfcrop} (a CASA\footnote{http://casa.nrao.edu}  task) for secondary calibrators flagging. Basic cross-calibration procedures such as delay, bandpass and gain calibrations were undertaken with \textsc{CASA}. Before self-calibration, we flagged target fields for RFI using \texttt{AOflagger} \citep{Offringa2010} with a strategy to only inspect the Stokes-Q amplitudes of the visibilities. The solution intervals obtained from calibration were then used to perform two iterations of self-calibration with  \texttt{Cubical} \citep{Kenyon2018}.

\subsection{\HI{} imaging}\label{subsec:imaging}
To obtain emission-only visibilities, we initially subtracted the continuum sky model generated after self-calibration. We then applied a third-order polynomial subtraction to the visibilities using \texttt{mstransform}. Additionally, each data cube has been Doppler-corrected to the barycentric reference frame and regridded with a channel resolution of 210 kHz (corresponding to a velocity resolution of $44.3$ \kms{} at $z=0$) and the same start frequency of 1309 MHz to ensure they have the same number of channels, enabling their combination into mosaics (see Section \ref{sec:mosaic}).

\begin{figure}
	\centering
	\includegraphics[width=\linewidth]{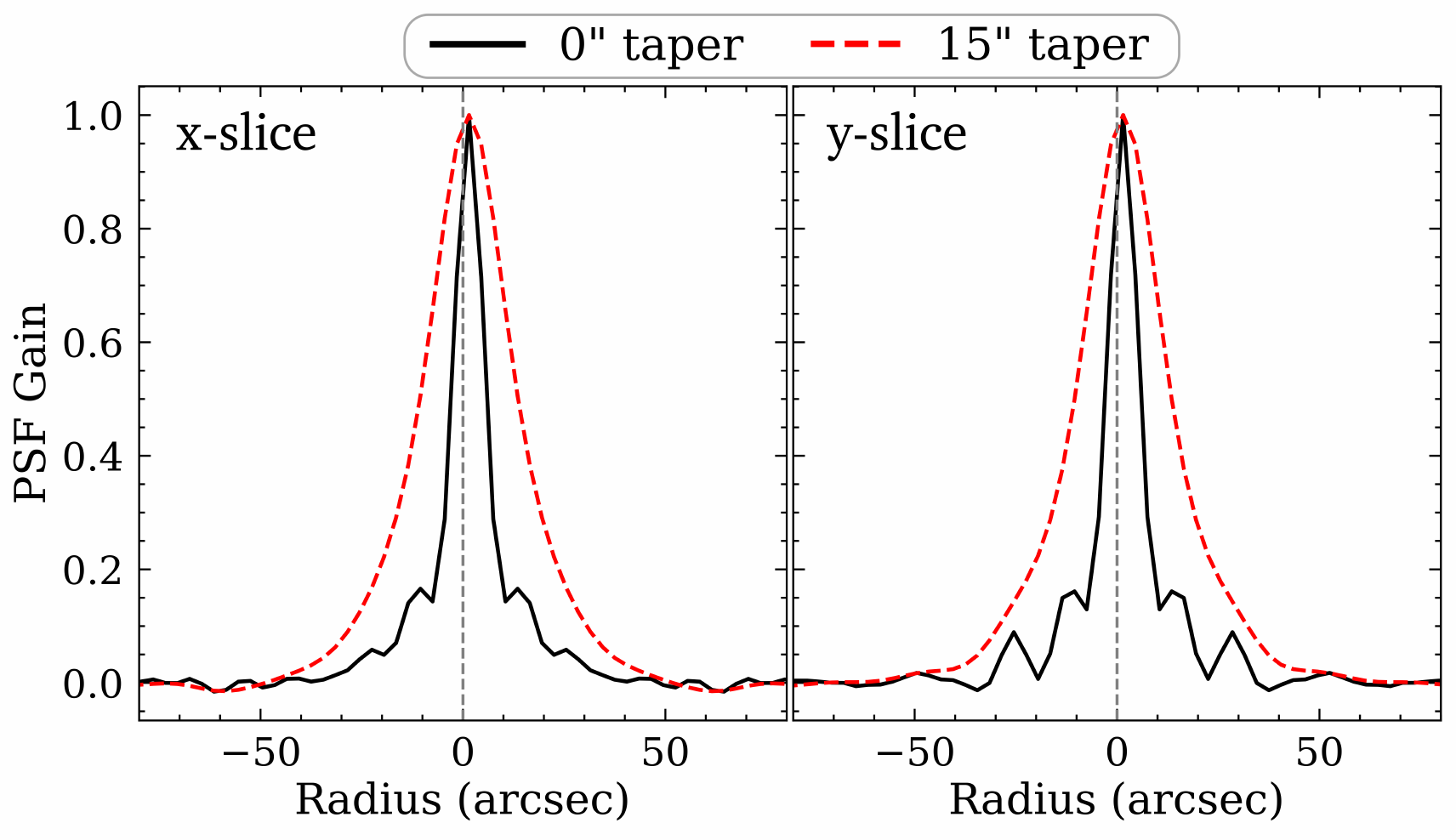}
	\centering
	\caption{A 2D slice of the dirty PSF along the x and y axes in a Vela$-$SMGPS field. The black solid line shows the PSF without taper, the red dashed line indicates the PSF after a $\ang{;;15}$ taper has been applied.}
	\label{fig:PSF}
	\vspace{-0.5em} %reduce space
	\end{figure}

\begin{figure}
	\centering
        \includegraphics[width=\linewidth]{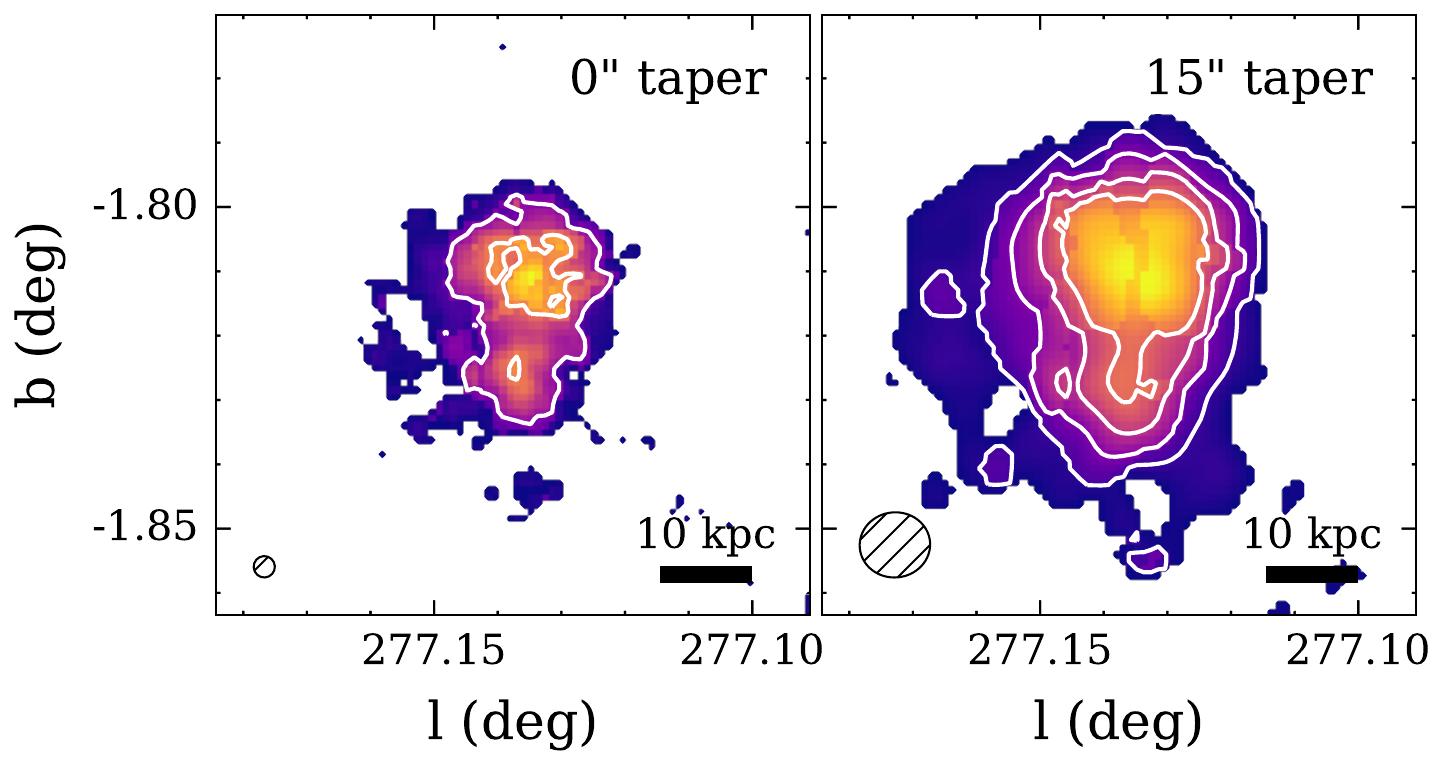}
	\centering
	\caption{ An example of the \hi{} distribution without a taper (left) and a $\ang{;;15}$ taper (right). PSFs with respective sizes of $\ang{;;12} \times \ang{;;11}$ and $\ang{;;30} \times \ang{;;27}$ are shown in the lower left corner of each image, while a 10 kpc scalebar is illustrated in the lower right corner.}
	\vspace{-0.5em} %reduce space
	\label{fig:HIZOA_taper}
\end{figure}

We produced 157 dirty \hi{} cubes of 570 channels each ($-1696 < V_{\rm hel} < 25515$ \kms{}) with a Briggs robust weighting parameter of $r = 0$, a pixel size of $\ang{;;3}$, and a taper of $\ang{;;15}$ using \textsc{WSClean} \citep{Offringa2014}. We introduced a taper of $\ang{;;15}$ to reduce the non-Gaussian pattern of the dirty Point Spread Function (PSF), i.e., smoothing the wings visible by the black solid lines in Fig. \ref{fig:PSF}, and improving the filling factor of extended low \hi{}-column densities sources to increase the image fidelity (e.g., see Fig. \ref{fig:HIZOA_taper}). The tapered dirty beams are on average $\ang{;;30} \times \ang{;;27}$ with a variation of about $\pm \ang{;;2}$ over the whole SMGPS survey, and approximately $\ang{;;29.7} \times \ang{;;26.3}$ ($\pm \ang{;;1}$) %, indicating a variation of ${\sim}$3.6\% 
in the 157 \hi{} cubes' beam sizes throughout Vela$-$SMGPS. We can therefore spatially resolve galaxies with \hi{} sizes larger than %$39 \times 35~(\pm 1.25)$ kpc or \hi{} masses $\log (M_{\mathrm{HI}}/\mathrm{M_{\odot}}) \gtrsim 9.62 \pm 0.03$ at the VSCL distance (${\sim} 18000$ \kms{}), 
${\sim}37 \times 33~(\pm 1.25)$ kpc or \hi{} masses of $\log (M_{\mathrm{HI}}/\mathrm{M_{\odot}}) \gtrsim 9.57 \pm 0.03$ at the VSCL distance, %(use 9.6 if round everything to 30 by 26)} 
based on the \hi\ size-mass relation \citep{Rajohnson2022}. 

\begin{figure*}
	\centering
	\includegraphics[width=\linewidth]{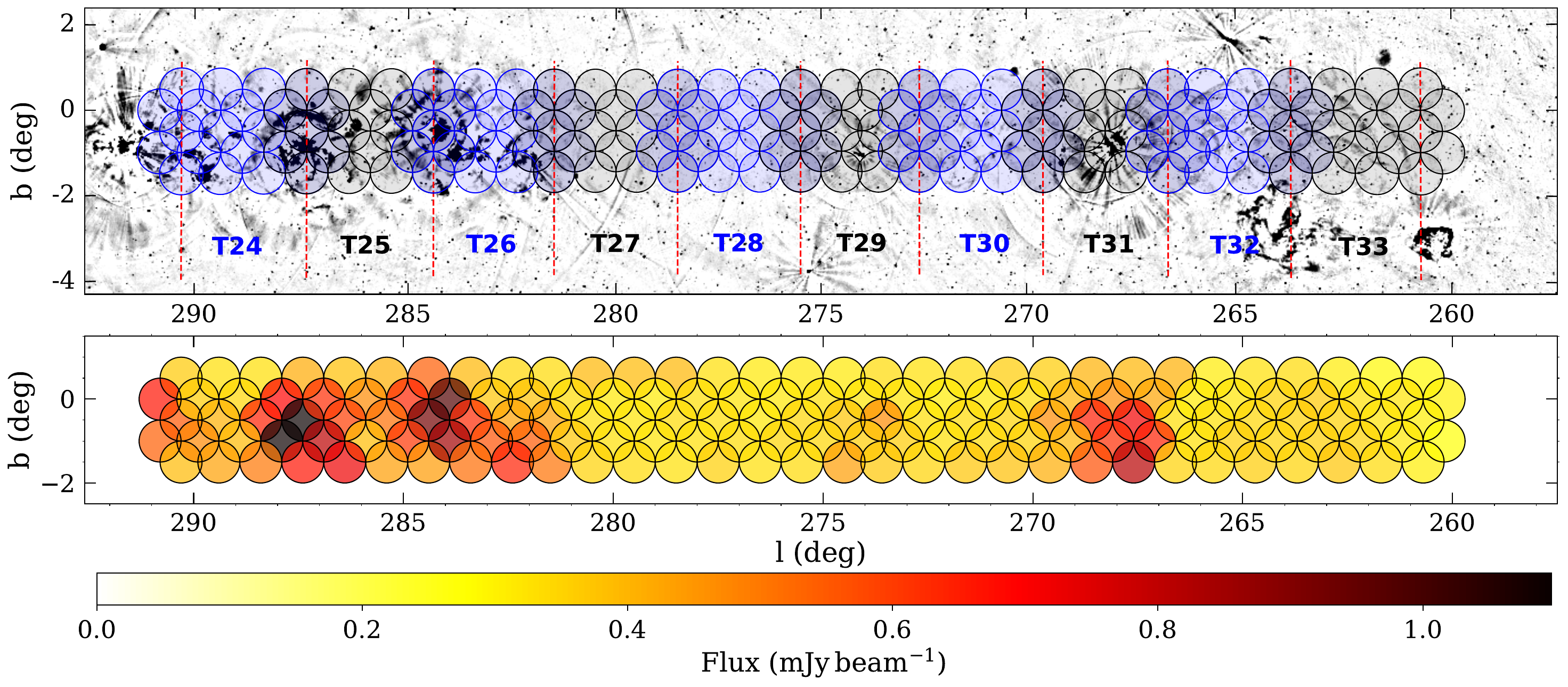}
	\centering
	\caption{ Pointing configuration of Vela$-$SMGPS ($260^{\circ} \leq \ell \leq 290^{\circ}$). \emph{Top panel:} distribution of the 10 mosaics (T24-T33) of 22 fields each in the VSCL region, overlayed on the SUMSS continuum map (grayscale). \emph{Bottom panel:} the corresponding rms levels of each field.}
	\label{fig:GPS_full}
	\end{figure*}

Figure \ref{fig:GPS_full} provides a visual representation of the pointing configuration for the 157 MeerKAT pointings in Vela$-$SMGPS. In the bottom panel, the color gradient, from yellow to dark red, illustrates the measured global rms noise values at the central region of each individual pointing. These values range from 0.28 to 1.09 \mJy{}. The fields with high rms noise (bottom panel) are closely related to the regions of extended continuum sources, e.g., $\ell \sim 283^{\circ} - 289^{\circ}$ and $267.5^{\circ}$ (see top panel). Fortunately, only a few fields are affected by these large and bright continuum sources. Non-contaminated areas, on average, exhibit an rms noise value of 0.42 \mJy{}. The necessary data quality for our science goals can be achieved by mosaicking the MeerKAT pointings in hexagonal patterns, as detailed in the next section. 

\subsection{Mosaicking strategy}\label{sec:mosaic}
	
We utilize the \textsc{Montage}\footnote{http://montage.ipac.caltech.edu/} software within CARACal with the \texttt{mosaic}\footnote{https://github.com/caracal-pipeline/MosaicQueen} worker parameter set to spectral type for the generation of the mosaics. To perform the assembling and re-gridding, we input the dirty cubes along with their ``Mauchian'' Primary Beam (PB) models (i.e., a cosine-squared power illumination as seen in \citealt{Mauch2020}) in a common folder. Fields of $0.8^{\circ}$ %0\fdg8
radius, corresponding to a normalized PB gain of 0.2, are used for mosaicking the PB-corrected dirty beams. This enhances the sensitivity and uniformity of the survey area. 

The restoring beams mentioned in Section \ref{subsec:imaging} vary from cube to cube and channel to channel. Therefore, the mosaic PSF is determined as the median value from the PSFs of each input cube in CARACal. We divide the Vela$-$SMGPS survey into 10 overlapping mosaics, named T24 to T33, each containing 22 fields. This results in each mosaic having its unique beam size. For example, mosaic T24 has a PSF of $\ang{;;28.9} \times \ang{;;25.0}$ and T33 has $\ang{;;30.5} \times \ang{;;26.3}$. These beam sizes are used to estimate the flux for the detections they contain. The varying beam areas per channel and per cube in the mosaics exhibit a scatter ranging between 3\% to 6\% compared to the median mosaic beam area. This scatter, however, is negligible when compared to the statistical uncertainty of the fluxes (see Section \ref{sec:ancillary}).\\
One mosaic covers an area of the sky slightly over $5 \times 3.5$ deg$^2$, producing a nearly uniform data cube of approximately $3.5 \times 3$ deg$^2$. Each mosaic centre is offset by $\Delta \ell = 3^{\circ}$. 
This configuration ensures that we do not miss possible extended sources at the edges of the mosaic. It allows an independent data quality assessment of the source finding algorithm and their derived \hi{} parameters from the overlapping area of $\Delta \ell \sim 1^{\circ}$ (see Section \ref{subsec:flux_verification}). \\

The CARACal run outputs three types of mosaics: the main mosaic containing the data and its corresponding weight and noise mosaics, all in fits format and projected in equatorial coordinates. To facilitate the analysis along the GP, we transformed the coordinates in Galactic coordinates using the CASA task \texttt{imregrid}, applied to the three mosaic files. 
The 10 final \hi{} mosaics are indicated in the top panel of Fig. \ref{fig:GPS_full} (red dashed lines), superimposed on the Sydney University Molonglo Sky Survey (SUMSS) Galactic continuum map \citep{Bock1999} in the top panel. The alternating blue and black MeerKAT pointings indicate the 22 pointings that conform one individual mosaic.

Due to the hexagonal configuration of the MeerKAT pointings, the rms after mosaicking is consistently lower. The 10 \hi{} mosaics have rms values ranging from 0.29 to 0.56 \mJy{}, with a mean rms noise of 0.39 \mJy{}. Assuming a 200 \kms{} linewidth galaxy and a mean global rms of 0.39 \mJy{}, we can achieve a $5\sigma$ \hi{} mass detection limit of 2.75$\times 10^9$ \msun\ at the VSCL distance ($V_\mathrm{hel} \sim 18 000$ \kms{}, $D \sim 257$ Mpc). Table \ref{tab:mosaic} lists the properties of the 10 Vela$-$SMGPS mosaics.

\begin{table}
    %\small
	\centering
	\caption{Mosaicking parameters of Vela$-$SMGPS.}
		\label{tab:mosaic}
		\begin{tabular}{@{} l c @{}}
		    \hline
			\hline
			Mosaicking property & Value \\
			\hline
			Number of fields per mosaic & 22 \\
			Number of mosaics & 10 (T24 to T33) \\
                Galactic longitude & $260^{\circ} \leq \ell \leq 290^{\circ}$ \\
                Galactic latitude & $-2^{\circ} \leq b \leq 1^{\circ}$ \\
                Total sky coverage & 90 deg$^2$ \\
			Sky coverage per mosaic & $\sim 5 \times 3.5$ deg$^2$ \\
			-- with uniform sensitivity & $3.5 \times 3$ deg$^2$ \\
			Velocity resolution &	44.3 \kms{} at $z = 0$ \\
			Mosaic velocity range & $ -1696 < cz < 25515$ \kms{} \\
			\hi{} cube weighting & Briggs \textit{robust} = 0 \\
			Pixel size & $\ang{;;3} \times \ang{;;3}$ \\
                \hi{} cube size &  $2200 \times 2200$ pixels \\
                Beam size ($\ang{;;15}$ tapering) & $\ang{;;29.7} \times \ang{;;26.3} ~ (\pm \ang{;;1})$ \\ 
			Mean global rms noise per channel &	0.39 \mJy{} \\
                Mosaics rms range &	$0.29-0.56$ \mJy{} \\
			\hline
		\end{tabular}
	\end{table}
	
\section{Source finding and verification}\label{sec:sofia}

The following section details the algorithms used to arrive at a reliable list of \hi{} detections. As a first step, we apply the automated source finder SoFiA-2 \citep{Westmeier2021} for source identification after optimizing its parameters, as discussed in Section \ref{subsec:sofia}. We discuss a visual verification and classification method in Section \ref{subsec:classification} to differentiate between solid detections and possible ones. In Section \ref{subsec:flux_verification}, a search for duplicates in the overlapping regions of the mosaics is performed to check for internal consistency, and a comparison with HIZOA is conducted to assess the quality of the measured fluxes.

\subsection{\HI{} source finding}\label{subsec:sofia}
It is not realistic to perform source finding by eye for such large surveys. Hence, we used the \hi\ Source Finding Application SoFiA-2\footnote{https://github.com/SoFiA-Admin/SoFiA-2} \citep{Westmeier2021}, an automated source finding and parameterization algorithm to identify galaxy candidates based on their \hi{} emission. Due to continuum residuals, the rms noise level from pointing to pointing is not uniform (see bottom panel of Fig. \ref{fig:GPS_full}). It is therefore crucial to optimize the parameter settings for SoFiA that performs reliably on all mosaics in finding the galaxy candidates, without resulting in a high number of false detections.\\ 

\begin{figure}
	\centering
	\includegraphics[width=\linewidth]{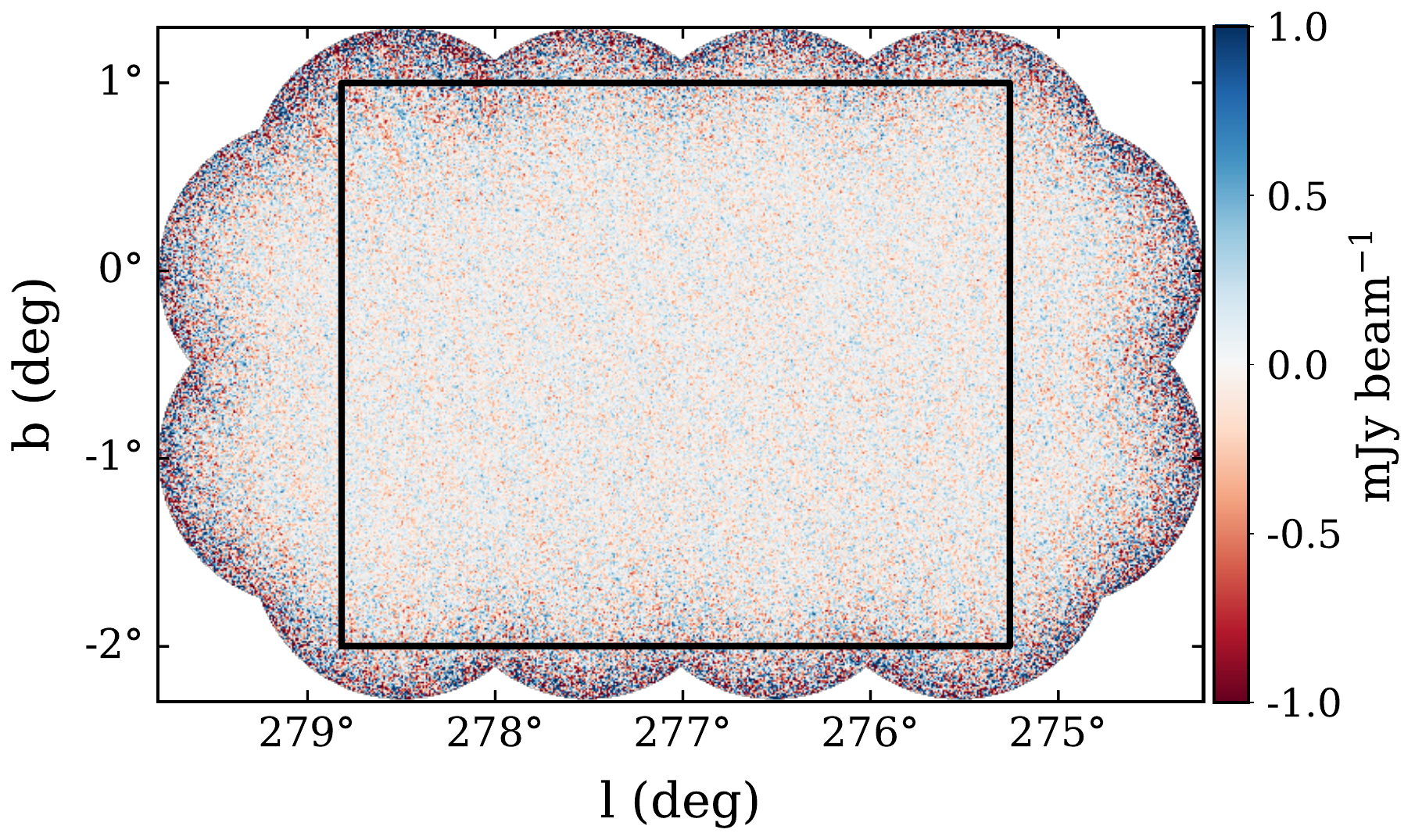}
	\centering
	\caption{ A random example showing spatial noise variation for a central channel map (rms = 0.29 \mJy{}) in a Vela$-$SMGPS mosaic. The black rectangle outlines the SoFiA source-finding region.}
	\label{fig:T28_SF}
	\end{figure}

\noindent
Our final source finding strategy is summarized below:
\begin{itemize}
    \item[(i)]  We first defined the search window for the input mosaic cube. Our three-dimensional search window was set to the central area of $3.5 \times 3$ deg$^2$ (${\sim}4200 \times 3600 \times 517$ pixels) where the rms is mostly uniform. Figure \ref{fig:T28_SF} displays a sample mosaic, illustrating that the spatial noise variation remains small within the search area outlined by the black box. The source finding velocity range was from channel 9 to 525, i.e., $250 \leq V_\mathrm{hel} \leq 25000$ \kms{}.
    \item[(ii)] We normalized the noise level across the mosaic by adding the noise mosaic as an additional input to the pipeline. We used the `local' scaling noise mode of SoFiA to correct for local noise variations within both the spectral and spatial domains of the data. This approach differs slightly from the one employed by \cite{Steyn2023_paper}, who used the `spectral' scaling noise mode, relying solely on spectral domain correction. 
    We chose a grid size of 301 pixels on the xy-axis and 15 channels on the spectral axis for this noise scaling step.
    \item[(iii)] The source finding step itself is performed with the `Smooth and Clip' (S+C) finder algorithm, specifically designed for blind source detection. It smoothes the mosaic with a set of spatial Gaussian filters and spectral boxcar kernels. We set these to (0, 7, 15) pixels, and (0, 135, 225, and 315) \kms, respectively. After each iteration of the smoothing process, the local rms noise is measured. SoFiA then clips all pixels below the source finding threshold of $4\sigma$. The remaining pixels are stored in a source mask and, if connected pixels satisfy the specified linking radius set in the parameter file, they are merged into individual sources.
    \item[(iv)] Each detected source undergoes a reliability assessment using the method outlined in \cite{Serra2012}. To be considered a valid detection and included in the SoFiA output catalog, a minimum SoFiA source-finding reliability threshold of 90\% is set.
    \item[(v)] To ensure that sources near the detection threshold of the survey were not overlooked due to stochastic noise, we conducted a second run of SoFiA with a source finding threshold of 3.5$\sigma$. Both resulting SoFiA catalogs were cross-matched using \texttt{Cross\_matching}\footnote{https://github.com/TrystanScottLambert/Cross\_Matching}, duplicates were removed, and unique detections from each run were retained. After a verification process discussed in the next section, we obtained a final list of detections which also allowed us to reject false positives.
\end{itemize}

\noindent
The final parameter settings, which encompass all these procedures, can be found in the SoFiA parameter file provided in Appendix \ref{app-A}. The outlined approach was applied to all 10 mosaics.

\begin{figure*}
	%\centering
	\subfigure{\includegraphics[width=\linewidth]{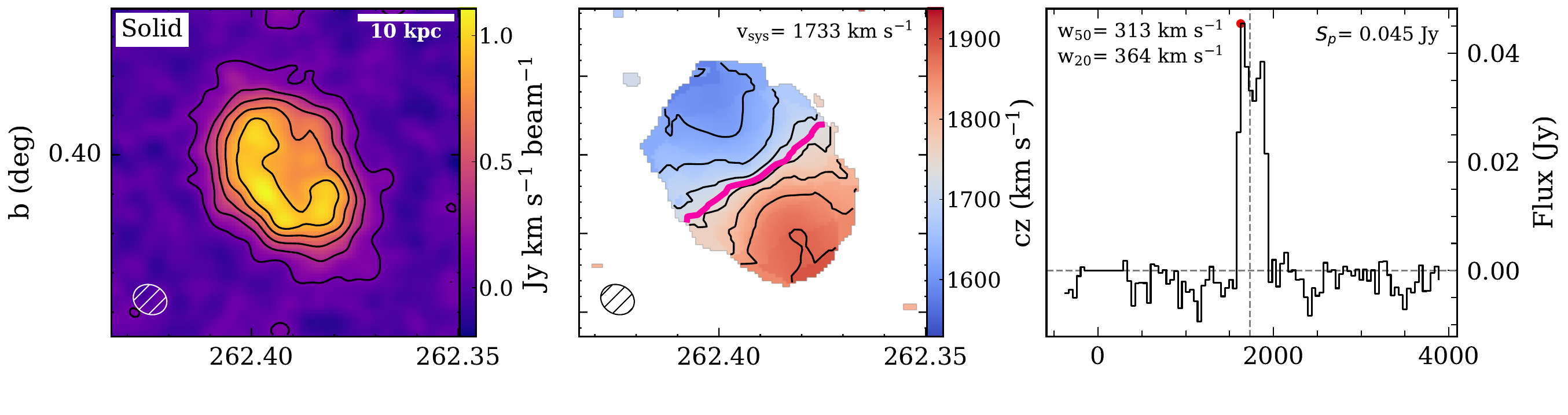}}\vspace{-2.5em}
        \subfigure{\includegraphics[width=1.01\linewidth]{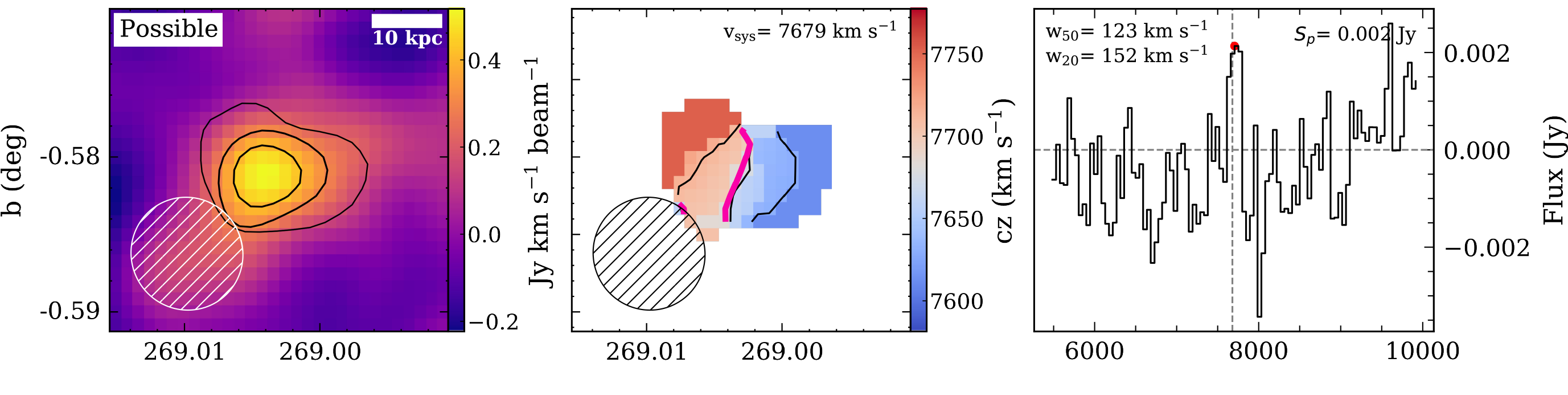}}\vspace{-2.0em} 
	\subfigure{\includegraphics[width=1.005\linewidth]{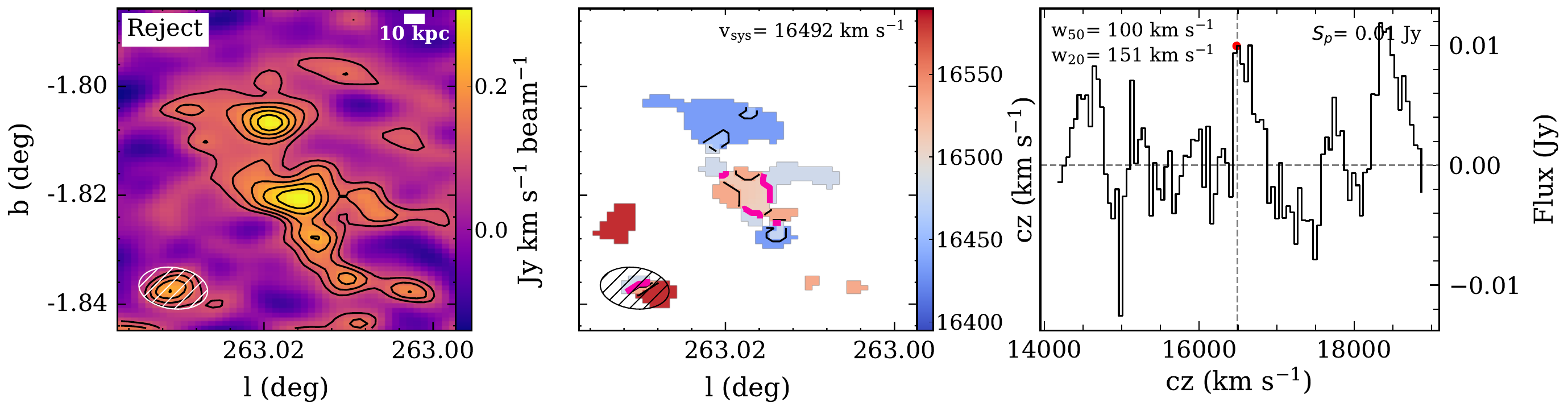}}
	\centering
	\caption{ The three visual classifications of SoFiA detected sources: ``Solid'', ``Possible'', and ``Reject'' (from top to bottom). \emph{Left panel:} Total flux (mom-0) map in Galactic coordinates with a color bar, a 10 kpc scale bar at the top, and the synthesized beam (white hatched ellipse) at the bottom corner. \emph{Middle panel:} Masked mom-1 map in Galactic coordinates with a color bar revealing velocity fields. The heliocentric velocity is given in the top right corner and indicated as a magenta contour. \emph{Right panel:} Global \hi\ profile with a dashed vertical line indicating the heliocentric velocity of the galaxy; measured linewidths at 50\% and 20\% of the peak flux density are presented in the top left corner, with the peak flux density value denoted by the red point shown in the top right corner.}
	\label{fig:classification}
	\end{figure*}

\subsection{Source classification}\label{subsec:classification}
For the verification process, a Python script was created to automatically generate total flux density (mom-0), velocity field (mom-1) maps, and the global \hi\ profiles for each detection listed in the SoFiA catalog, following the format shown in Fig. \ref{fig:classification}.

To generate moment maps and global \hi{} profiles, we extracted cubelets with a spectral length of 10 MHz and dimensions defined by the bounding box results obtained from the SoFiA catalog around each \hi{} detection. Additionally, we added extra padding of $\ang{;;30}$ to $\ang{;1}$ in both longitude and latitude depending on the redshift of the detection to account for the noise outside the source. 
To compute mom-0 and mom-1 maps, we selected the cubelet channels both before and after the systemic velocity provided by SoFiA, each with widths equal to half of the linewidth at the 20\% peak flux density ($w_{20}$) and an additional channel on both sides. A 3$\sigma$ clipping threshold was therefore applied, where $\sigma$ is the local rms derived from four emission-free areas around each detected \hi{} signal, used to mask the moment maps.

After the production of these data products, we performed visual verification simultaneously on the spectrum, unmasked mom-0 maps (to assess noise levels surrounding the source), and on masked mom-1 maps. Each potential detection was classified as `solid', `possible', or `reject' according to the following qualitative guidelines: 

To be considered a solid detection, firstly, the global \hi{} profile should exhibit a clear peak with a relatively flat baseline, and the signal should be well above the noise level. Additionally, the source should have a well-defined disc in the mom-0 map and a distinct velocity field with evident rotation over at least two channels in the mom-1 map.

Special attention is given to compact sources less extended than 1.5 times the beam size. Such sources typically have an integrated signal-to-noise ratio (SNR) of less than 6, indicating low completeness and reliability. Faint unresolved sources with SNR $\leq 6$ (see Eq. \ref{eq:snr}) were conservatively moved from the solid to the possible detection list as they were difficult to distinguish from noise peaks visually. Similarly, in cases of low-velocity gradient (especially if a single channel peak), sources with low SNR are moved to the possible detection list.

In cases where one of these criteria is not met, but the detection is still deemed to be real, it is classified as a possible detection. An example is shown in the second panel of Fig. \ref{fig:classification}, where despite an evident rotation spanning four channels, its extent is only about the size of the beam. Additionally, this detection presents an integrated SNR of approximately 5. The highest point in its spectrum is also not very clear due to the varying noise in the detection area. 

A detection is otherwise labeled as a reject. For instance, the existence of an extremely large source compared to its beam size (see bottom panel of Fig. \ref{fig:classification}), or a small dwarf, is not likely to be detectable in this survey at high redshift. This could also arise from a small noise peak (few pixels) that SoFiA misinterpreted as a detection, a source detected in more than 20 channels, or the identification of an imaging artifact caused by a residual continuum. The latter assessment was facilitated by an additional visual verification of each detection using CARTA\footnote{https://cartavis.github.io} -- an interactive Astronomy Visualization tool. The adjudication process involved multiple team members to ensure the robustness of the source verification. Figure \ref{fig:classification} displays typical examples of these three categories.

\begin{figure}
	\centering
	\includegraphics[width=\linewidth]{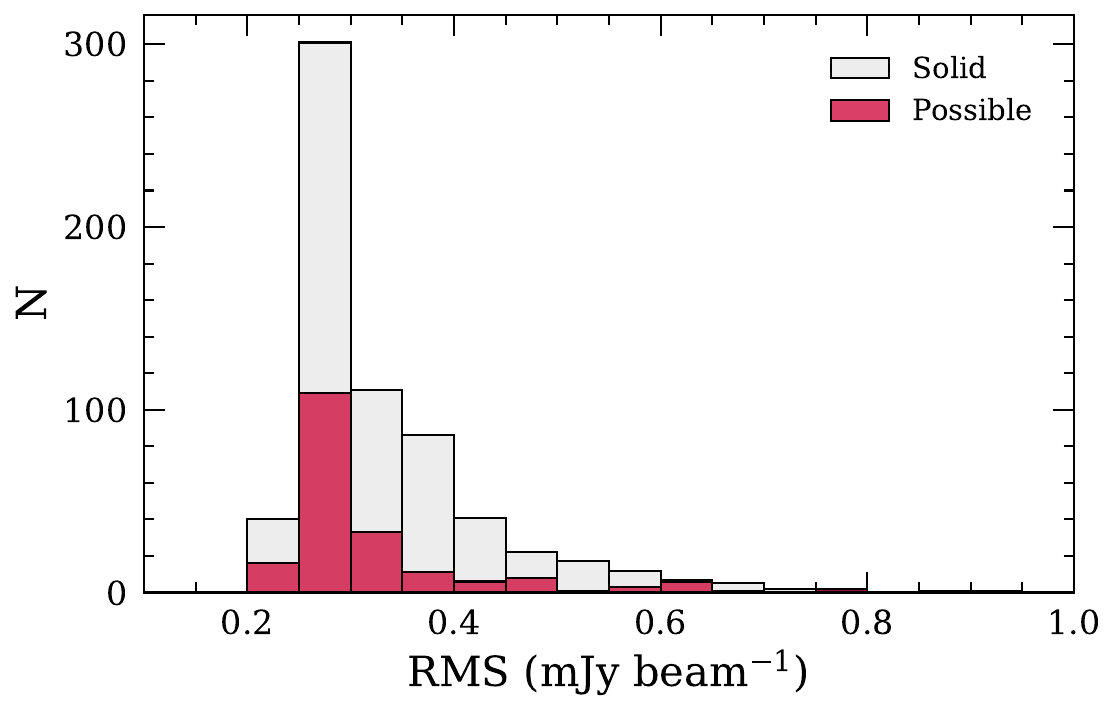}
	\centering
	\caption{Histogram depicting the local rms noise measured around each solid detection (shaded in grey) and possible detection (shaded in red).}
	\label{fig:rms}
	\end{figure}

We then cataloged a total of 843 detections, comprising 646 solids and 197 possible detections based on this classification scheme. Given that SoFiA provides an estimate of the reliability of each source as defined by \cite{Serra2012}, we used those individual reliability values to determine the estimated number of false detections present in the catalog. We estimate a value of 6.4\% for false detections among possible detections, while solid detections yield a false detection rate of just 1.4\%. Since multiple team members also inspected the moment maps visually, these estimates are likely to represent upper limits.\\
These detections were compiled into a Galaxy Atlas, which showcases their masked moment maps and \hi\ global profiles. The Galaxy Atlas is available at \url{https://doi.org/10.5281/zenodo.11160742}. Table \ref{tab:sofia} provides the number of detected sources per SoFiA run and per mosaic, their final classifications, as well as the average global and local rms variations per mosaic. The per-channel local rms noise of the 843 detections ranges from 0.22 to 0.94 \mJy{}, with an average of 0.33 \mJy{}. This is lower than the average global rms value of the mosaics and is consistent for solid and possible detections (see peak in Fig. \ref{fig:rms}). The solid and possible detections from each mosaic are combined into a final catalog (see Section \ref{sec:cat} for details) with flags according to their classifications.

\begin{table*}
    \small
	\centering
	\caption{Number of detected sources per mosaic after the two SoFiA runs, including cross-matching, elimination of duplicates between mosaics, and visual verification. The last column presents the measured global and local rms values determined around detections.}
		\label{tab:sofia}
		\begin{tabular}{@{} c | c c | c | c c | c c @{}}
		    \hline
			\hline
			Mosaic &  \multicolumn{2}{c |}{Thresholds} & Total retained & Solid & Possible & \multicolumn{2}{c}{Mean rms (\mJy{})}\\
			%\hline
				& $3.5\sigma$ &	$4\sigma$ & (w/o dupl.) &	 &  & Global & Local\\
			\hline
            T24 & 51 & 65 & 54 & 42 
            & 12 & 0.47 & 0.39 $\pm$ 0.12\\	
            T25 & 56 & 58 & 32 & 25 
            & 7 & 0.49 & 0.50 $\pm$ 0.12 \\
            T26 & 60 & 72 & 53 & 48 
            & 5 & 0.50 & 0.41 $\pm$ 0.11\\
            T27 & 137 &	155 & 116 & 97 
            &   19	&   0.30	&   0.32   $\pm$	0.09\\
            T28 & 108 &	116 & 102 & 65 
            &	37 &   0.29	&   0.28   $\pm$ 0.04 \\
            T29 & 98 &	111 & 91 & 70 
            &	21	& 0.31	&   0.29   $\pm$ 0.06\\
            T30 &	120	& 134 & 114 & 	92 
            &	22	 &   0.30	&   0.31    $\pm$ 0.08 \\
            T31 &	119	& 107 & 75 & 52 
            & 23 &   0.56    &	0.38    $\pm$ 0.09 \\
            T32 &	131	& 125 & 103 & 72 
            & 31 &   0.31	&   0.31    $\pm$ 0.08 \\
            T33 &	138	& 139 & 103 & 83 
            &	20 &   0.33    &	0.30   $\pm$ 0.07\\
			\hline
			\multicolumn{3}{c |}{Total} & \textbf{843} 
            & \textbf{646}
            & \textbf{197} 
            & & \\ \cline{1-6}
		\end{tabular}
	\end{table*}

\subsection{Parameterization}\label{subsec:parameterization}

The SoFiA output list provides source parameters in native pixels and flux units. To convert them into physical parameters, such as centroid positions (in pixels) into Galactic coordinates ($\ell$ and $b$ in degrees), and to enable SoFiA to adjust the integrated fluxes ($S_{\rm int}$) by the beam solid angle, we set SoFiA \texttt{parameter.wcs} and \texttt{parameter.physical} to \texttt{True}. 

We first convert the SoFiA output parameters. The heliocentric systemic velocity $V_{\rm hel}$ is determined from the central frequency of the \hi{} profile relative to the optical definition using the equation:
\begin{equation}
    V_\mathrm{hel} = c \left( \frac{\nu_0}{\nu} - 1\right)  = cz,
\end{equation}
where \emph{c} is the speed of light, $\nu_0$ is the rest-frequency of the 21 cm line, and $z$ is the observed barycentric redshift. No corrections have been applied to $V_\mathrm{hel}$ for the barycentre of the Local Group ($V_{\mathrm{LG}}$). 

In a few isolated cases, where the detection was composed of one or more companions, the SoFiA central frequency has been found slightly off-center. We then determined the systemic velocity as follows:
\begin{equation}
    V_\mathrm{hel} = \rm 0.25(v_{20}^{A} + v_{50}^{A} + v_{20}^{R} + v_{50}^{R}),
\end{equation}
where $\rm v_{20}$ and $\rm v_{50}$ are the corresponding velocities measured at the 20\% and 50\% levels of the peak flux densities (see \citealt{Ramatsoku2016}). A and R indicate the approaching and receding sides of the profile respectively. \\

We not only determined the systemic velocities but also calculated the linewidths and integrated fluxes of the detections. It is important to emphasize that the conversions were conducted using the source rest frame velocity convention, following the method outlined in \cite{Meyer2017}. 

To convert ($w^{\nu}_{20}$, $w^{\nu}_{50}$) from their default units in Hz to \kms{}, we use the equation:
\begin{equation}\label{eq:width}
    w = \frac{c(1+z)}{\nu_0} w^{\nu}.
\end{equation}
When the signal-to-noise ratio is low, SoFiA could output an under- or overestimate, where the values of $w_{20}$ and $w_{50}$ are erroneously identical, contrary to what is expected in the spectrum. In such cases, we re-calculated the linewidths as:
\begin{equation}
    w_{20} = \mathrm{v_{20}^{R} - v_{20}^{A} ~ and ~}
    w_{50} = \rm v_{50}^{R} - v_{50}^{A}.
\end{equation}
The 20\% and 50\% levels are measured from the first crossings of the peak flux density. The differences between the SoFiA linewidths and the re-calculated ones are on the order of half to one channel width.
Additionally, the resulting linewidths have been corrected for instrumental broadening according to the method outlined in \cite{For2021} and \cite{Springob2005}:
\begin{equation}
    w^c = \frac{c(1+z)}{\nu_0} \sqrt{(w^\nu)^2 - (\Delta \nu)^2},
\end{equation}
where $\Delta \nu = 210$ kHz is the observed frequency width. Using Eq. \ref{eq:width}, this can be further simplified to:\\
\begin{equation}\label{eq:w50_corr}
     w^c = \sqrt{w^2 - (\Delta V)^2} .\\
\end{equation}

Moreover, the default output from SoFiA for the integrated flux $S^\nu$ is in Jy Hz. To determine the \hi{} mass in Solar masses (\msun{}), we directly apply equation 48 from \cite{Meyer2017}:
\begin{equation}\label{eq:hi_mass}
    \left( \frac{M_{\rm HI}}{\rm M_{\odot}} \right) = 49.7 \, \left( \frac{D}{\rm Mpc} \right)^2 \, \left( \frac{S^\nu}{\rm Jy\,Hz} \right),
\end{equation}
where $D = V_{\rm hel}/H_0$ represents the Hubble distance in Mpc.
For comparison purposes in the following sections, we convert the integrated fluxes ($S^{\nu}$) from Jy Hz to Jy \kms{} ($S^{V}$), using the source rest frame convention:
\begin{equation} \label{eq:int_flux}
    \begin{aligned}
    S_{\rm int} = \left( \frac{S^V}{\mathrm{Jy~km~s^{-1}}} \right) = \frac{c(1+z)}{\nu_0} \, \left( \frac{S^\nu}{\rm Jy~Hz} \right).
    \end{aligned}
\end{equation}

\subsection{Quality assessment}\label{subsec:flux_verification}
%\noindent
\subsubsection{Internal quality assessment}\label{sec:internal}

\begin{figure}
	\centering
	\includegraphics[width=0.85\linewidth]{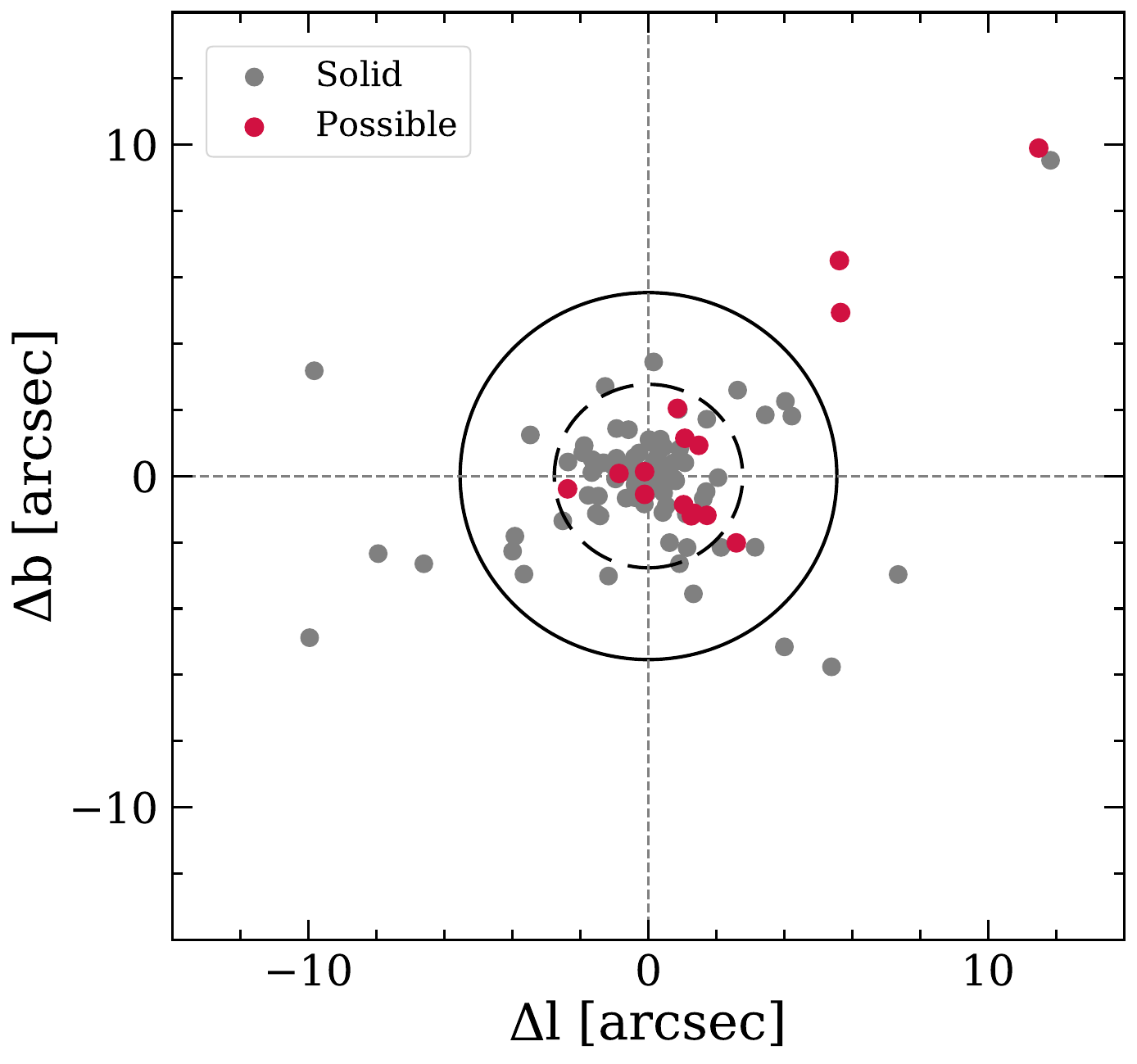}
	\centering
	\caption{ Internal coordinate offsets for detections found on overlapping regions of adjacent mosaics. Grey dots illustrate solid detections while red ones designate possible detections. The inner (dashed) and outer (solid) circles delimit the one and two sigmas standard deviations of $\ang{;;2.77}$ and $\ang{;;5.54}$, respectively.}
	\label{fig:coordprec}
	\end{figure}

\begin{figure}
	\centering
	\includegraphics[width=\linewidth]{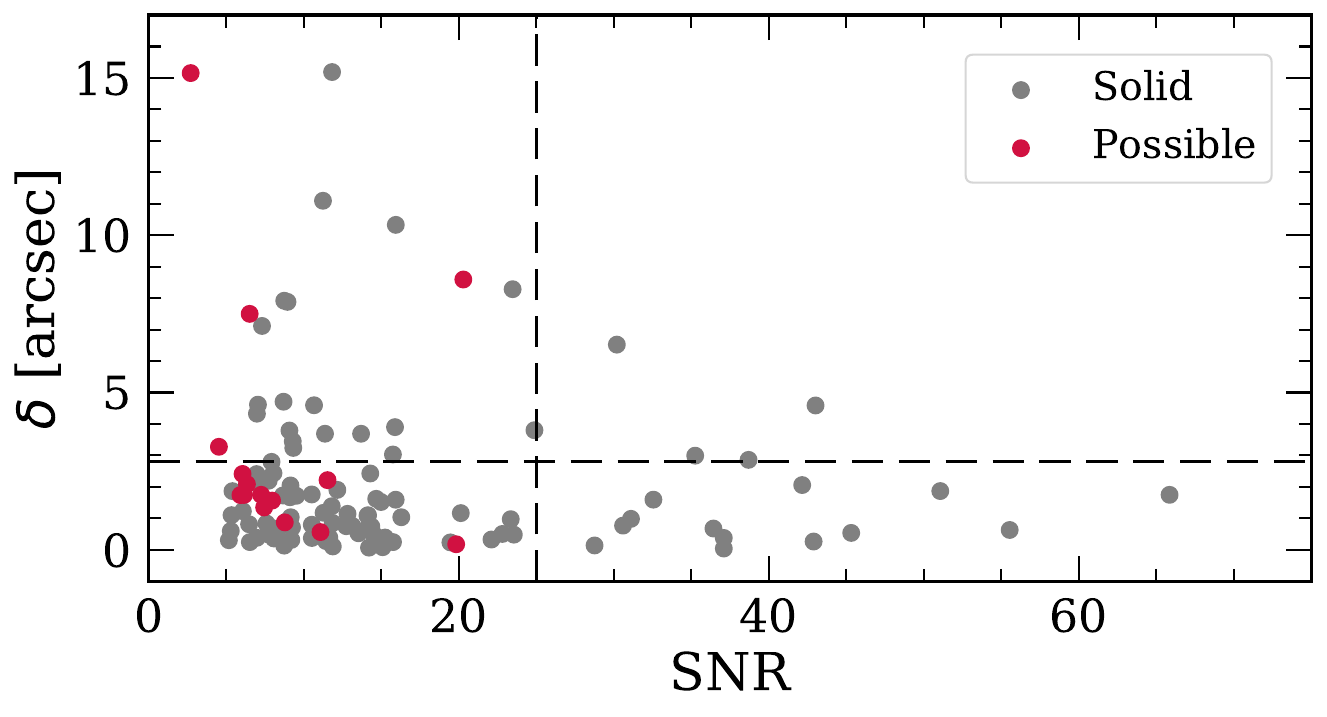}
	\centering
	\caption{  Vela$-$SMGPS internal coordinate offsets as a function of integrated signal-to-noise ratio (SNR). Red and grey dots represent possible and solid detections. Horizontal and vertical dashed lines mark the coordinate precision 1$\sigma$ standard deviation, and an SNR of 25, respectively.}
	\label{fig:coordprec_snr}
	\end{figure}

We first conducted an internal consistency check for sources located within the overlapping regions of the mosaics to ensure consistency in source-finding, and to assess internal data quality. A source from an overlapping region was identified in adjacent mosaics consisting of a slightly different configuration of pointings. We first investigated the positional agreement as demonstrated in Fig. \ref{fig:coordprec}. A mean coordinate precision of $\ang{;;2.16}$ with a $1\sigma$ standard deviation of $\ang{;;2.77}$ was achieved -- less than one pixel size. No difference in positional offset between possible and solid detections was observed. We then examined the positional offsets as a function of the integrated signal-to-noise ratio (SNR; see Fig. \ref{fig:coordprec_snr}). The integrated signal-to-noise ratio is defined as: 
\begin{equation}\label{eq:snr}
    \mathrm{SNR} = \frac{\sum S_{i}}{\sigma \sqrt{N_{\mathrm{pix}} \Omega_{\rm PSF}}},
\end{equation}
following the definition from Equation 4 in the SoFiA user manual\footnote{https://gitlab.com/SoFiA-Admin/SoFiA-2/-/wikis/home\#resources-and-documentation}. In this equation, $\sum S_i$ refers to the summation of flux densities within the source mask, $\sigma$ is the local rms noise measured around the detection, $N_{\rm pix}$ denotes the total count of spatial and spectral pixels involved in the detection, and $\Omega_{\rm PSF}$ indicates the beam solid angle measured in pixels. 

Possible detections have on average a lower SNR compared to solid ones, while the $2\sigma$-outliers (outside black solid circle in Fig. \ref{fig:coordprec}) all have SNR $< 25$. The coordinate separation is seen to decrease with increasing SNR as shown in Fig. \ref{fig:coordprec_snr}. 

The absolute differences in integrated fluxes, linewidths, and systemic velocities of the overlapping sources (i.e., the same object detected in two different mosaics) exhibit mean absolute errors and standard deviations in the range of $0.17 \pm 0.12$ Jy \kms{}, $22 \pm 27$ \kms{}, and $10 \pm 12$ \kms{}, respectively. These values are consistent with their errors, and uncertainties in linewidths and velocities are well below the coarse channel resolution. However, for sources with very low SNR and detections situated close to the mosaic edges, the masking threshold may vary if the rms values differ in the two mosaics, potentially leading to larger flux uncertainties.  

 \subsubsection{Ancillary data comparison}\label{sec:ancillary}
 
In the next step, we quantify the accuracy of Vela$-$SMGPS \hi{} parameters. 
Despite the high extinction in the inner ZOA, we utilized the NASA/IPAC Extragalactic Database (NED)\footnote{http://ned.ipac.caltech.edu}, and the Vizier\footnote{https://vizier.cds.unistra.fr/viz-bin/VizieR} Catalogue access tool to search for Optical, IR, and \hi{} counterparts within our survey region. The search radius was determined in quadrature to the positional uncertainty of Vela$-$SMGPS (as discussed in Section \ref{sec:internal}) and the typical coordinate precision of the catalog used for comparison. For instance, we conducted searches within a radius of $\ang{;;10}$ from each galaxy's position for OPT/IR counterparts, and ${\sim}\ang{;4}$ for \hi{} counterparts, considering that HIZOA galaxies \citep{Staveley2016} are with few exceptions the only \hi{} counterparts available at these latitudes. \\

Out of 843 detections, 57 detections (${\sim}7$\%) have likely IR and/or \hi{} counterparts in the existing literature. Among these 57, the distribution of IR counterparts is as follows: 7 (12\%) have IRAS counterparts, 40 (70\%) are associated with 2MASS/2MASX counterparts, 50 (88\%) are identified with WISE counterparts, and 17 (30\%) have IRSF counterparts. The median coordinate separation between the \hi{} position and the potential IR counterpart is approximately $\ang{;;3.3}$. 

In terms of \hi{} counterparts, our survey area encompasses 39 HIZOA detections (68\% of 57), with 6 of them also detected in HIPASS \citep{Meyer2004}. Vela$-$SMGPS successfully retrieved all HIZOA detections. Interestingly, we recovered the most \hi{}-massive spiral galaxy hidden behind the Milky Way, identified as HIZOA J0836-43 \citep[see][]{Kraan2005_hizoa}, named SMGPS-HI J083650-433738 in Vela$-$SMGPS (displayed in the top left panel of Fig. \ref{fig:hiZOA_unique1}). This galaxy is the most \hi{}-massive object within the entire HIPASS volume, and underwent subsequent follow-up observations with the Australia Telescope Compact Array (ATCA; \citealt{Donley2006}). This optically obscured galaxy was found to be a Luminous Infrared Galaxy (LIRG) with an actively star-forming disk spanning over 50 kpc \citep{Cluver2010}. 

We used both the ATCA and HIZOA data for this \hi{} massive galaxy to examine the consistency in our positions, fluxes, velocities, and linewidths. We observe a positional difference of $\, \ang{;;9}$ and $\ang{;;23}$, respectively. While the ATCA and HIZOA measurements report integrated \hi{} flux densities of $14.5 \pm 0.7$ Jy \kms{} and $14.38 \pm 1.69$ Jy \kms{}, our analysis using Vela$-$SMGPS yields a value of $15.15 \pm 1.59$ Jy \kms{}. All three measurements fall within the range of expected errors, indicating excellent agreement with each other. The slightly higher flux value of J083650 in Vela-SMGPS may be attributed to its position near the mosaic edge. We observed higher noise levels and baseline irregularities in this area, which could have influenced this case's flux measurement. Moreover, the systemic velocities and linewidths exhibit exceptional agreement with the HIZOA results after correcting the linewidths for instrumental broadening. Both show differences of 8 \kms{} in the velocities and $w_{20}$, while only 1 \kms{} for $w_{50}$. When compared to ATCA data, there is a 10 \kms{} difference in velocities, and larger offsets of 30 \kms{} and 15 \kms{} are observed in the $w_{50}$ and $w_{20}$ values, respectively. 

The comparison with the sole interferometric dataset within our survey region provides consistent measurements. However, the statistical analysis is limited to only one galaxy. Hence, we utilize the HIZOA dataset for an independent comparison of \hi{} parameters, given its unique status as the only source with redshift information in our multi-wavelength counterpart search. \\

\begin{figure}
	\centering
	\includegraphics[width=0.8\linewidth]{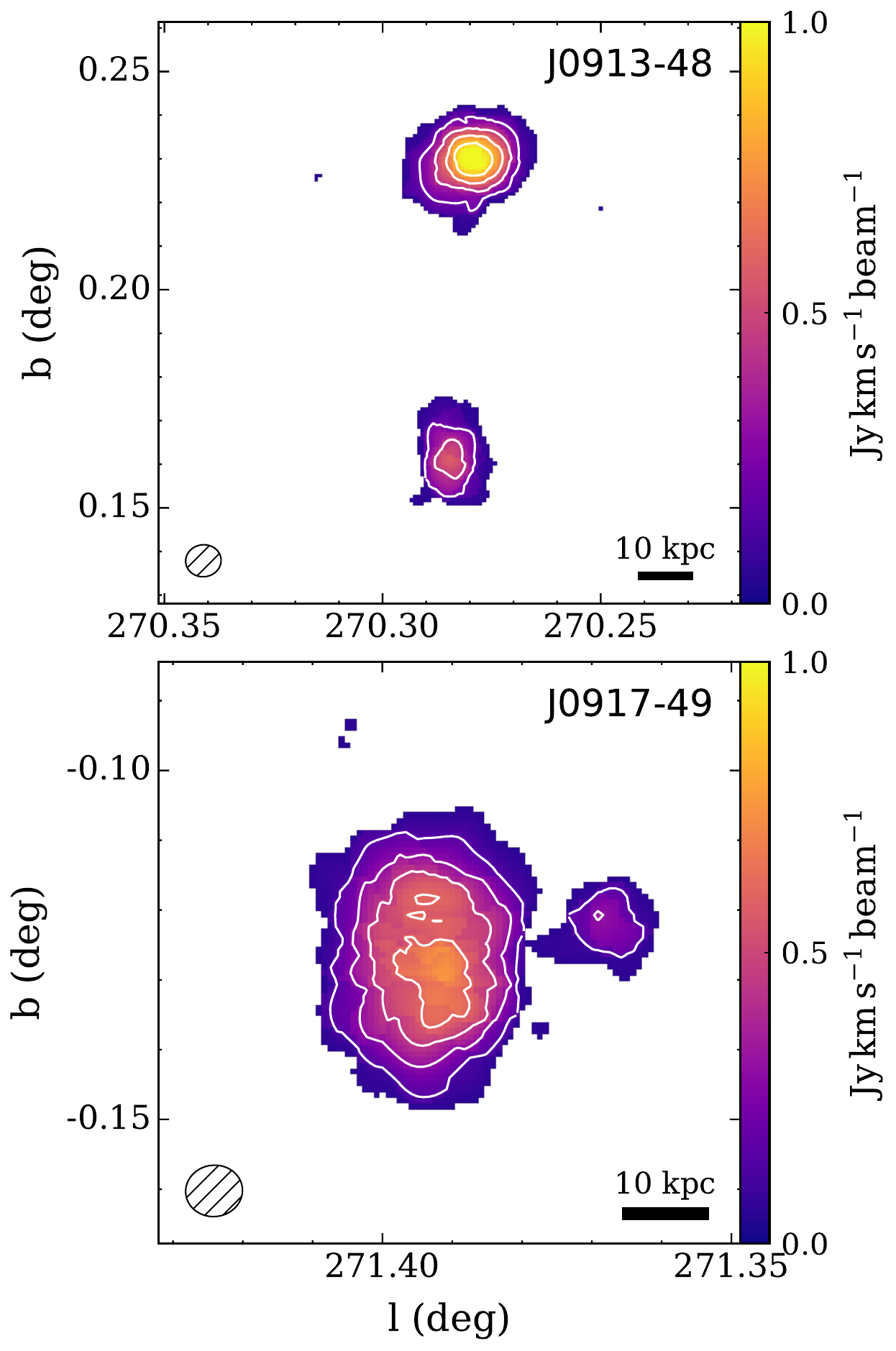}
	\centering
	\caption{ Moment-0 \hi{} emission maps for two HIZOA sources (J0913-48, and J0917-49) reveal them to consist of several counterparts in Vela$-$SMGPS. A 10 kpc scale bar and the beam size are shown in the right and left bottom corners of each image.}
	\label{fig:HIZOA_mom}
	\end{figure}

The 39 HIZOA galaxies are identified out to 12000 \kms{}. The Vela$-$SMGPS dataset comprises 471 detections within this velocity range, marking a noteworthy twelve-fold increase in the galaxy count. The mean positional offset is ${\sim}\ang{;2}$, a reasonable value given the large ${\sim}\ang{;15.5}$ beam size of HIZOA. This is also consistent with findings between HIZOA/HIPASS and other data sets \citep[e.g.,][]{Staveley2016}. Six of the HIZOA detections were resolved into several distinct \hi{} sources due to the higher resolution of the MeerKAT interferometric data. Figure \ref{fig:HIZOA_mom} presents an illustration of two examples, showcasing HIZOA J0913-48 blended into two distinct sources, and HIZOA J0917-49 with a satellite companion as seen in their \hi\ distributions. Moment maps of the 39 HIZOA detections as observed by Vela$-$SMGPS are shown in Appendix \ref{app-B}. Furthermore, 19 of these HIZOA detections have likely WISE counterparts.\\

\begin{figure}
    \centering
    \includegraphics[width=0.9\linewidth]{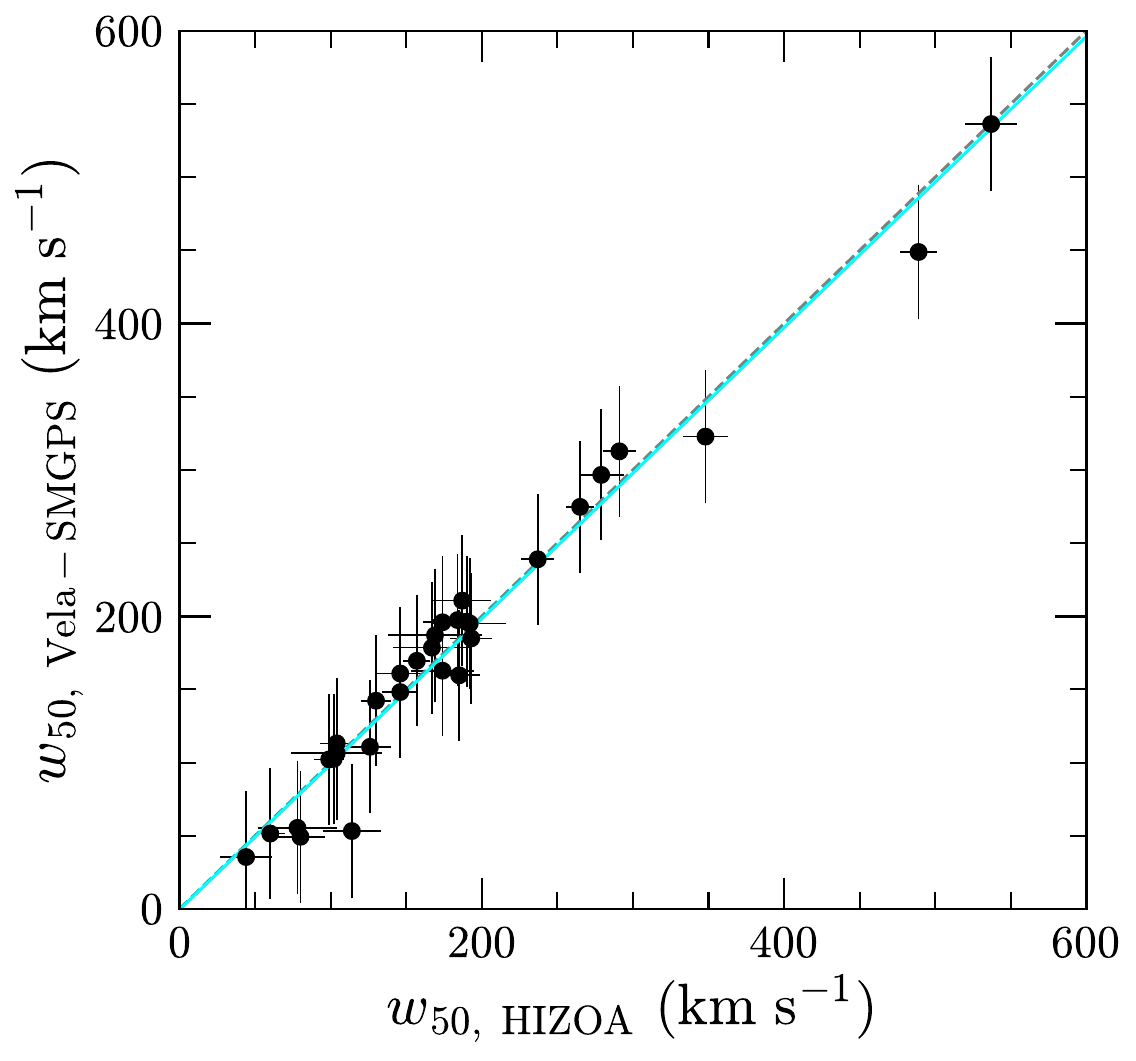}
    \caption{ Comparison of \hi{} linewidths $w_{50}$ among the 31 HIZOA detections within the Vela$-$SMGPS survey. The cyan line represents the best-fit linear regression}.
    \label{fig:comp_w50}
\end{figure}

\begin{figure}
    \centering
    \includegraphics[width=0.9\linewidth]{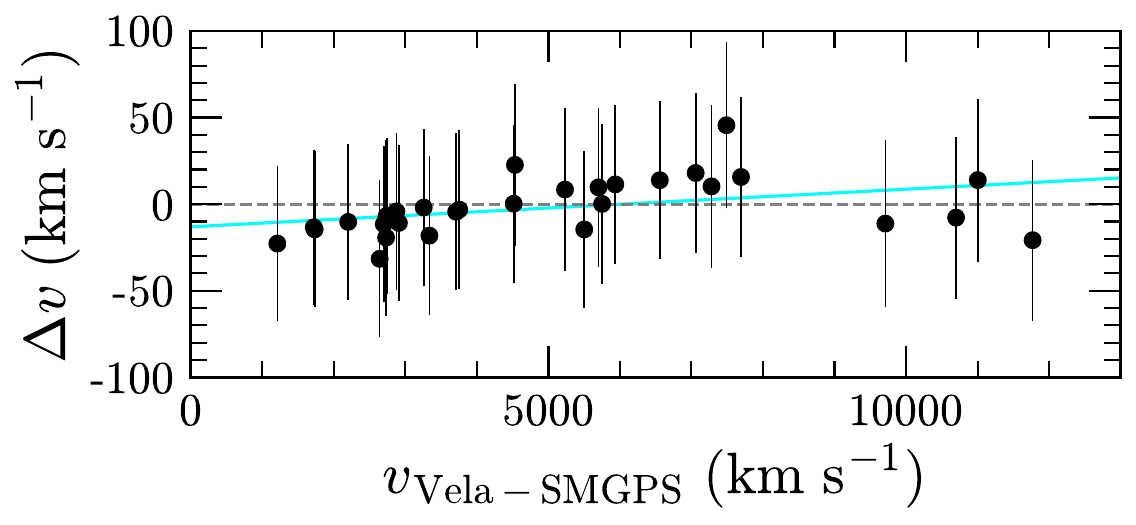}
    \caption{Comparison of Heliocentric Velocities from Vela$-$SMGPS and the difference $V_{\rm Vela-SMGPS} - V_{\rm HIZOA}$.}
    \label{fig:comp_vopt}
\end{figure}

\begin{figure}
    \centering
    \includegraphics[width=0.9\linewidth]{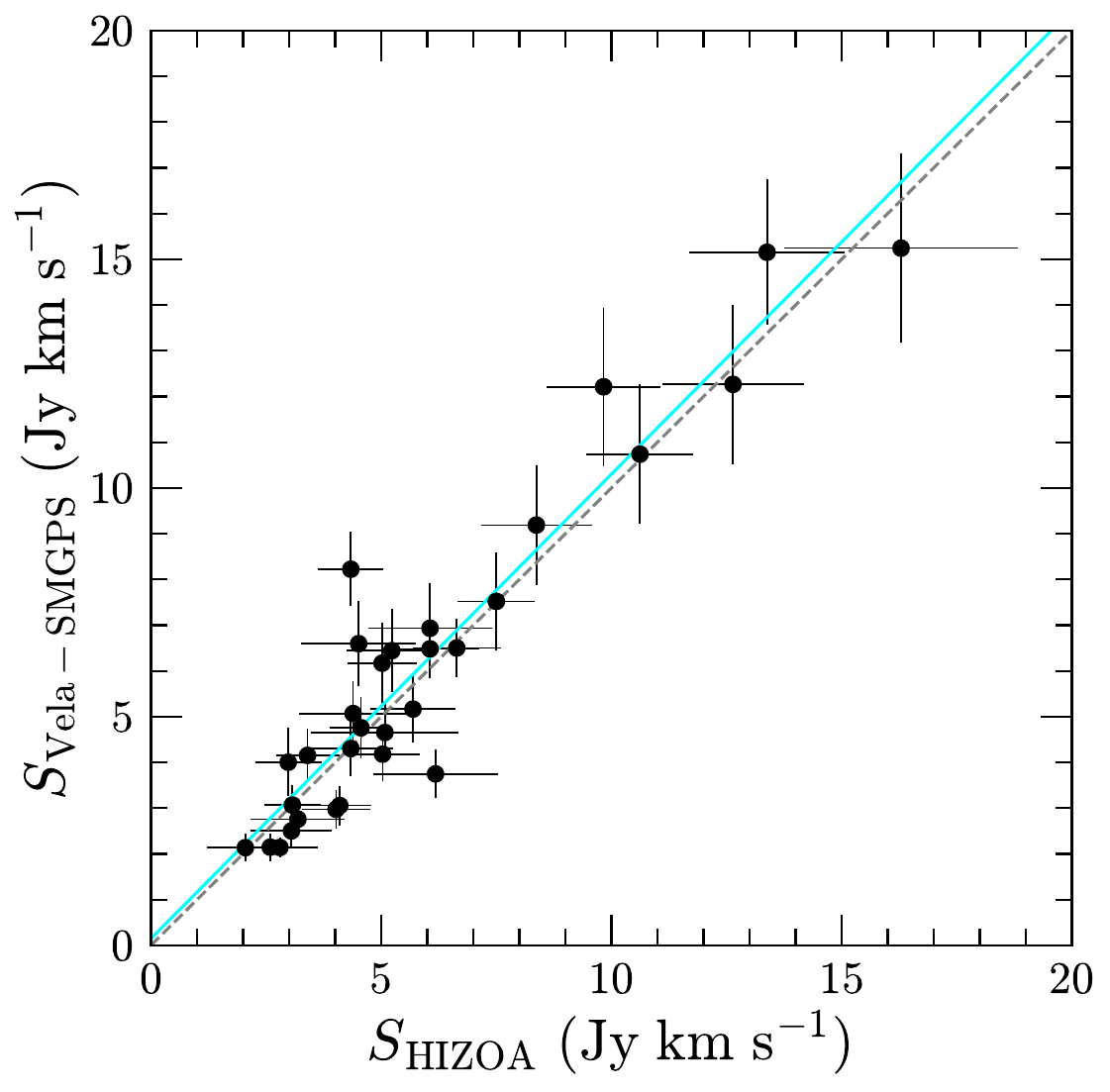}
    \caption{ Comparison of the integrated fluxes of 31 galaxies in common
with HIZOA. The cyan line illustrates the best-fit linear regression, closely
following the one-to-one relation indicated by the grey dashed line.}
    \label{fig:comp_int_flux}
\end{figure}

Figures \ref{fig:comp_w50}, \ref{fig:comp_vopt}, and \ref{fig:comp_int_flux} display the comparison between the linewidths, systemic velocities, and integrated fluxes of the Vela$-$SMGPS detections and their corresponding HIZOA counterparts. Out of the 39 detections, we excluded detections that are conformed of multiple counterparts or close to a continuum residual artifact, leaving 31 detections for comparison. 

The 50\% peak flux density linewidths show a nearly indistinguishable deviation from a one-to-one relation (black dashed line), with a mean error of approximately 15 \kms{} and a standard deviation of 12 \kms{}. A formal fit yields a regression with a slope of $0.994 \pm 0.067$ (cyan line) reaching an R$^2$ value of 0.97 at a 95\% confidence interval. Similarly, the one-to-one linear fit for Vela$-$SMGPS and HIZOA heliocentric velocities also exhibits a slope of almost unity ($1.002 \pm 0.002$). We chose to plot their differences, $\Delta v = V_{\rm Vela-SMGPS} - V_{\rm HIZOA}$, for a more comprehensive comparison. The results show a mean error of $15 \pm 10$ \kms{}. Given the coarse SMGPS channel width (44.3 \kms{}) and the (smoothed) HIZOA spectral resolution (27 \kms{}), this achievement is remarkable, affirming the high quality of the SMGPS calibration and \hi{} parameters.

We furthermore verified the agreement in integrated fluxes, as illustrated in Fig. \ref{fig:comp_int_flux}. To derive the integrated flux error $\varepsilon_s$, we adopt the method used by \cite{Ramatsoku2016}. We selected four emission-free regions located at the corners of the cubelet by dividing each cubelet into a $3 \times 3$ grid of the same size as the source mask. $\varepsilon_s$ is then computed as the mean of the standard deviations of the measured integrated fluxes within these boxes. We observe typical flux uncertainties of ${\sim}32\%, 19\%, 14\%$, and 10\%  for \hi{} masses below $10^8, 10^9, 10^{10}$ and above $10^{10}~\rm M_{\odot}$, respectively. The measured slope, nearly equal to unity ($1.015 \pm 0.135$), along with R$^2 = 0.89$ at a 95\% confidence interval, underlines the consistency between SMGPS and HIZOA flux measurements with less than 2\% offset to unity. Overall, the parameters of the comparison sample show excellent agreement with our measurements.\\

 \begin{figure*}
	\centering
        \subfigure{\includegraphics[width=0.49\textwidth]{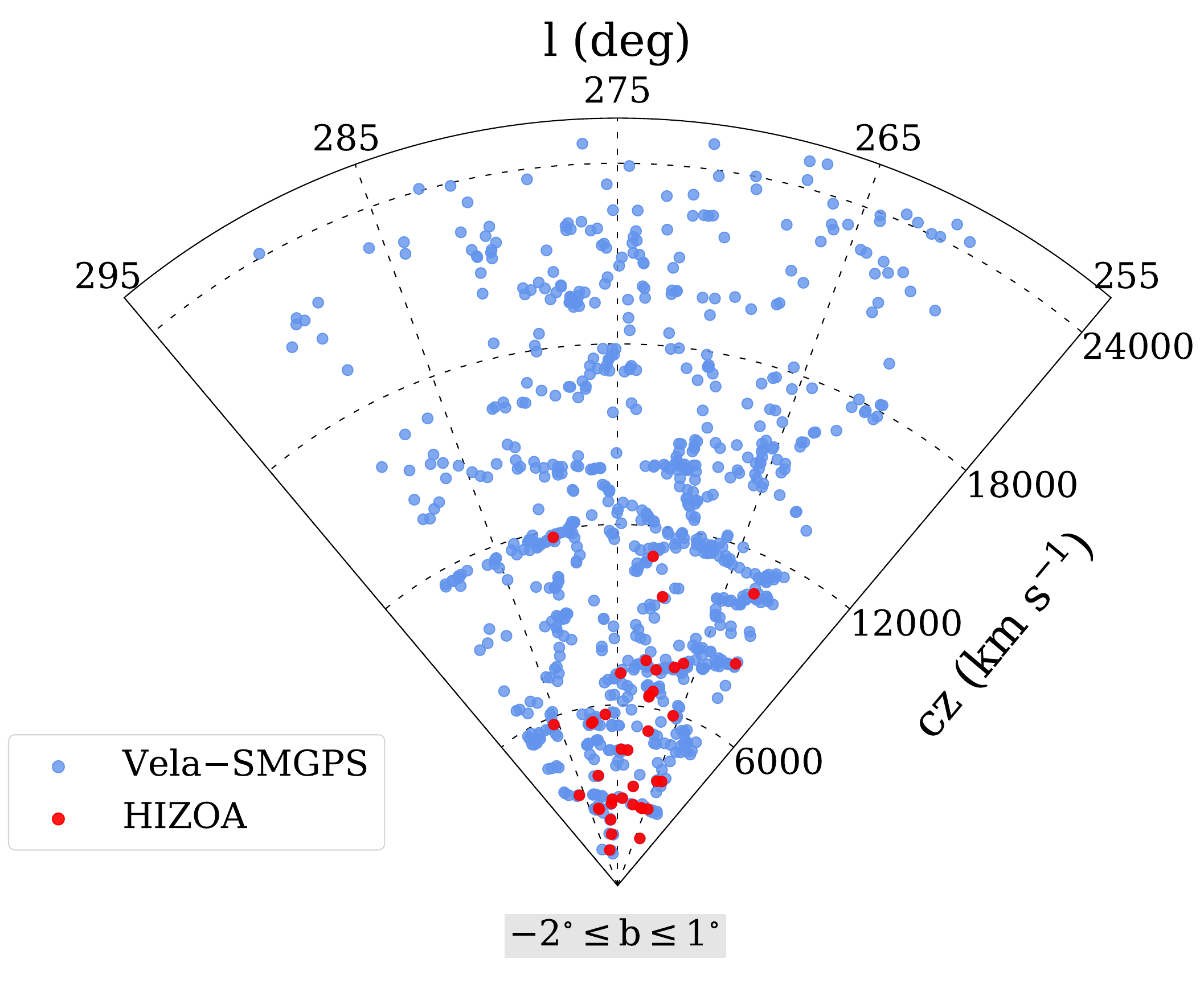}}
	\subfigure{\includegraphics[width=0.38\textwidth]{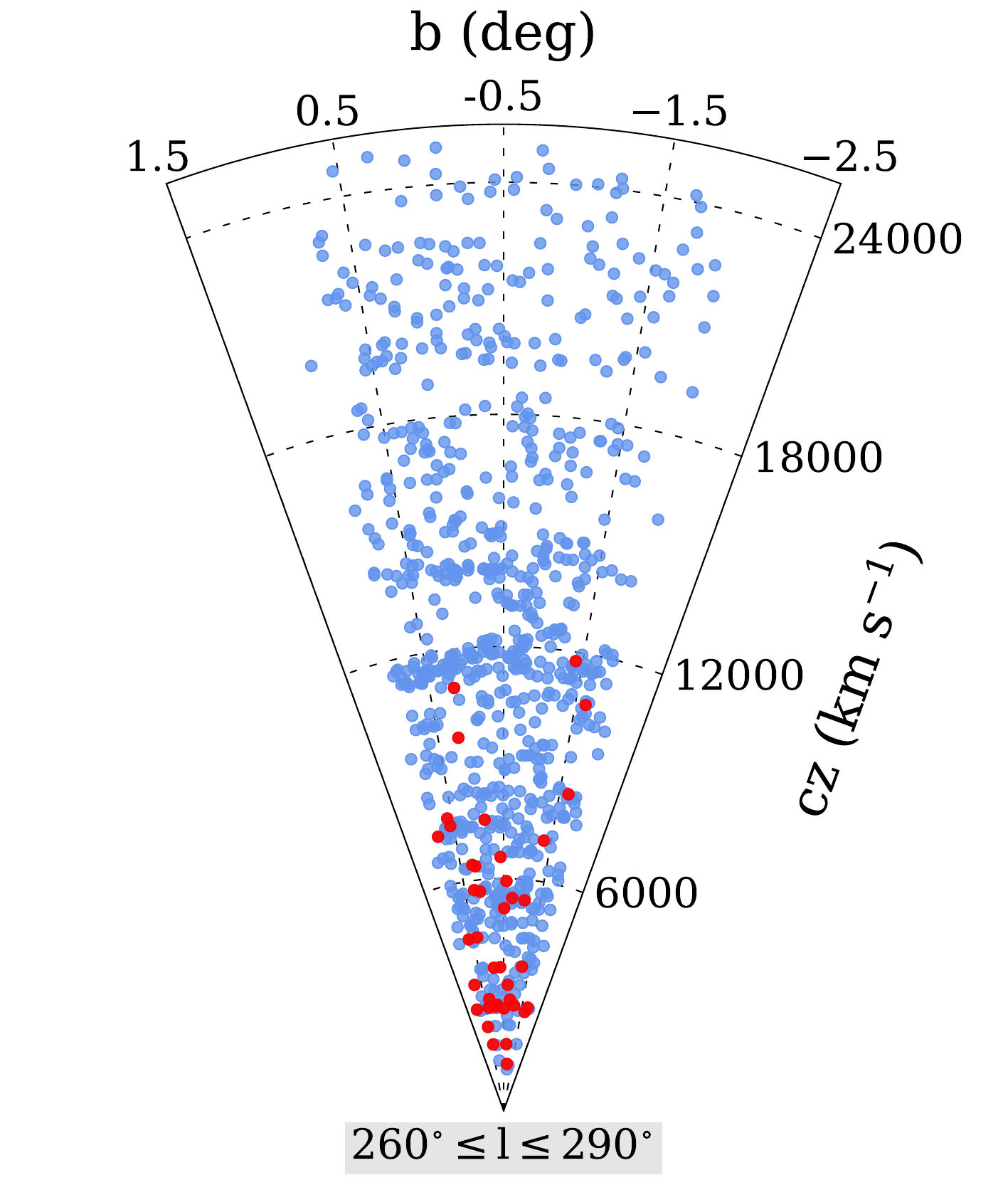}}
    
	\centering
	\caption{ Redshift wedge diagrams illustrating the distribution of \hi{}-detected galaxies up to $V_{\mathrm{hel}} < 25000$ \kms{} in Vela$-$SMGPS (blue dots) and HIZOA (red dots). The left panel depicts a longitude wedge plot covering the latitude range $-2^{\circ} \leq b \leq 1^{\circ}$, while the right panel displays a latitude wedge plot spanning the longitude range $260^{\circ} \leq \ell \leq 290^{\circ}$.}
	\label{fig:wedge_hizoa}
	\end{figure*}
	
\section{Results}\label{sec:cat}
The resulting Vela$-$SMGPS catalog contains 843 \hi\ detections. Across the full velocity range ($V_{\rm hel} < 25000 $ \kms{}), this corresponds to an average of ${\sim}$9.37 galaxies per deg$^2$. Around $93$\% of the galaxies have no optical or IR counterpart. Apart from the 39 HIZOA galaxies $+1$ ATCA (massive galaxy), no other redshift measurements exist. Of the 843 detections, 187 galaxies have velocities within $19 500 \pm 3500$ \kms\ and might be associated with the Vela overdensity. Of these, 76 galaxies lie within the velocity range of $16000-19000$ \kms{}, and 111 galaxies within the $19000-23000$ \kms{} range, which conform to the two walls of the VSCL. The \hi{} parameters of the 843 \hi{} detections are compiled into a catalog, which is presented in Appendix \ref{app-D} as Table \ref{tab:cat}.\\

Figure \ref{fig:wedge_hizoa} illustrates the wedge diagrams in Galactic longitude and latitude of the Vela$-$SMGPS detections which demonstrate the impressive sensitivity of this survey. The newly detected \hi{} galaxies are indicated by the light blue dots, and the previously known HIZOA galaxies by red dots. The majority of the galaxies below 4000 \kms{} were also found by HIZOA with its sensitivity of 6 \mJy{}. At higher velocities, the HIZOA dataset predominantly captures the more massive spiral galaxies, but the uncovered structure closely aligns with the general shape (filaments) delineated by Vela$-$SMGPS. A detailed discussion of the LSS uncovered by Vela$-$SMGPS will be presented in Section \ref{sec:lss}.\\

\subsection{Completeness}\label{sec:completeness}

\begin{figure}
	\centering
	\includegraphics[width=\linewidth]
 {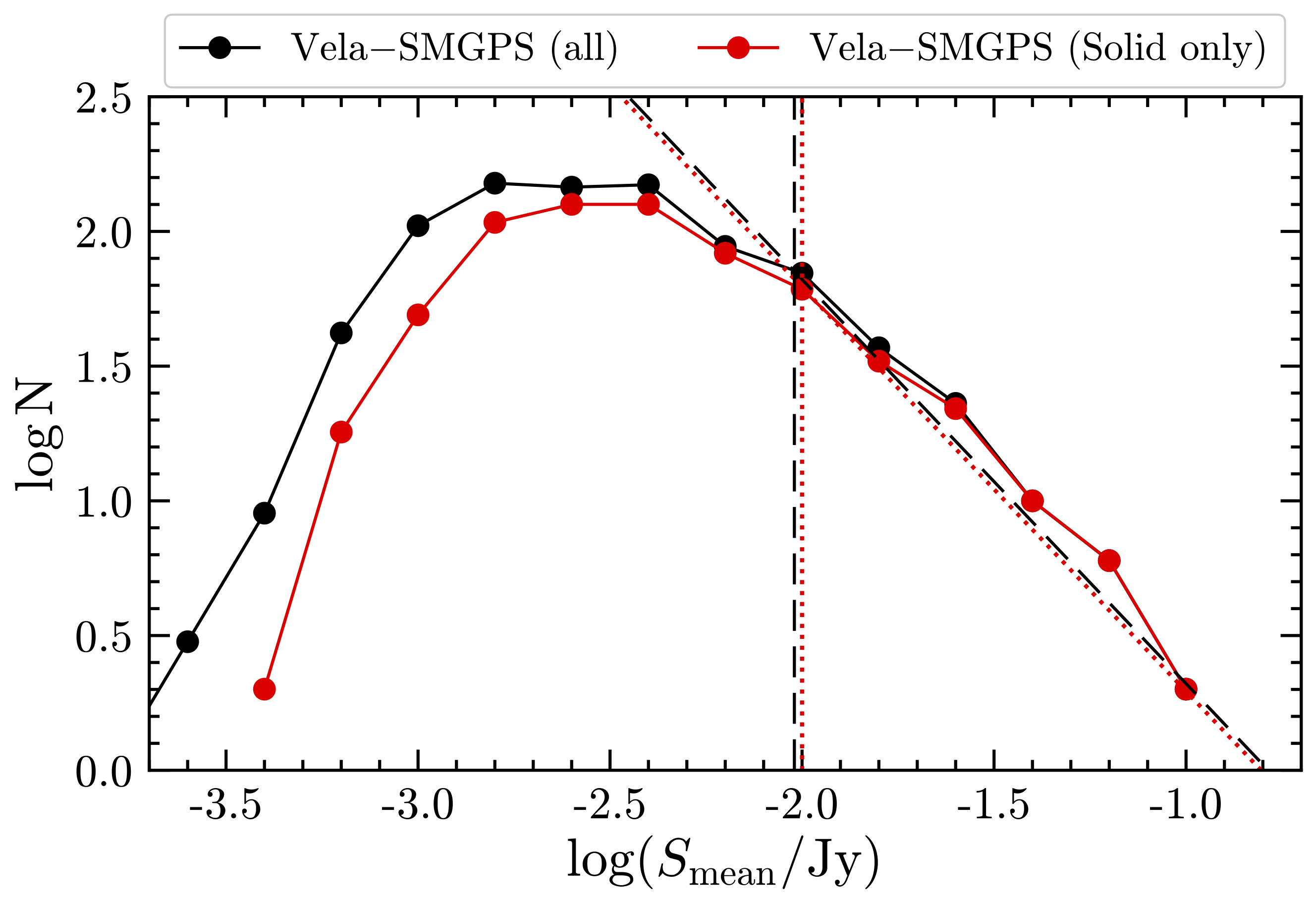}%{images/completeness_VSCL_GPS_simulation_v6.pdf}
	\centering
	\caption{Completeness assessment using a log-log plot of the mean \hi{} flux density $S_{\mathrm{mean}} = (S_{\mathrm{int}} / w_{50})$ for all 843 Vela$-$SMGPS detections in black and 646 solid detections exclusively in red. The black dashed and red dotted lines correspond to the expected $-3/2$ slope for a homogeneous distribution in both samples.}
	\label{fig:completeness}
	\end{figure}

One way to determine the completeness of a survey involves fitting a power law to the histogram of mean flux density, calculated as the integrated flux divided by the linewidth (cf. \citealt{Donley2005,McIntyre2015,Staveley2016}). The distribution is expected to follow a Euclidean power law, $N(S) \sim S_{\mathrm{mean}}^{-2.5}$, corresponding to a linear power law with a slope $-3/2$ in log-log space. The completeness threshold becomes apparent when the histogram distribution starts to deviate from this projected line. In Fig. \ref{fig:completeness}, we illustrate solid detections (red), as well as the sample including both solid and possible detections (black). We use $w_{50}$ rather than $w_{20}$ because $w_{20}$ is more influenced by noise variations. The difference in completeness is a mere 0.02 dex in the last bin, indicating that galaxies with $S_{\mathrm{mean}}$ exceeding 9.5 and 10 mJy are detected with high completeness in the full and solid samples, respectively. The similarity implies that possible detections are also quite reliable galaxy candidates. We therefore include both solid and possible detections in the further analysis. However, it is important to note that this method works best under the assumption of survey homogeneity and isotropy. A more rigorous completeness test \citep{Rauzy2001} will be conducted in a forthcoming paper.

\subsection{Simulations}\label{sec:simulations}

For a better interpretation of our results, especially in discerning the presence of overdensity peaks indicative of LSS, %\textcolor{green}{relative to a homogeneous distribution}, 
we compare our \hi\ data with a simulation. The latter, referred to as S45, utilizes a similar framework as the one discussed in \cite{Staveley2015} for the SKA-\hi{} science case. In this approach, constant \hi{} mass and velocity width functions are assumed across redshifts, using the bivariate stepwise maximum likelihood method outlined in \cite{Zwaan2005}. The simulation randomly distributes galaxies up to the observations sensitivity threshold (see also \citealt{Steyn2023_paper}), providing output \hi{} parameters such as peak flux density, integrated fluxes, \hi{} masses, linewidths, and redshifts, without any spatial structure. Here, S45 is tailored to the specifications of the SMGPS, i.e., a sensitivity of 0.45 \mJy{} to account for the average noise variation, the coarse channel widths of 44.3 \kms{}, and a survey area of 528 deg$^2$ similar to the full SMGPS survey.

As a result, S45 does not simulate LSS. Instead, it represents a close proxy for a homogeneous distribution. S45 predicts the number of galaxies (with SNR\,$>5$) as a function of redshift, distributed over an area equivalent to the full SMGPS. 
For the specification of the full SMGPS, it predicts 5591 galaxies in the volume delimited by $V_{\mathrm{hel}} < 25000$ \kms{}, respectively 953 galaxies for the Vela survey region (528 deg$^2/$90 deg$^2$). By comparing it to our observations, we can identify and quantify peaks and gaps indicative of overdensities and underdensities. While LSS in the data can be discerned visually, some structures may result from observational effects. Thus, using S45 enables a clear distinction between real structures and those influenced by selection effects, facilitating a more robust characterization of LSS through its comparison with a homogeneous galaxy distribution (see Section \ref{sec:overdens}).

\section{Vela$-$SMGPS \HI{} properties distribution}
\label{sec:properties}

\begin{figure}
	\centering
	\includegraphics[width=0.9\linewidth]{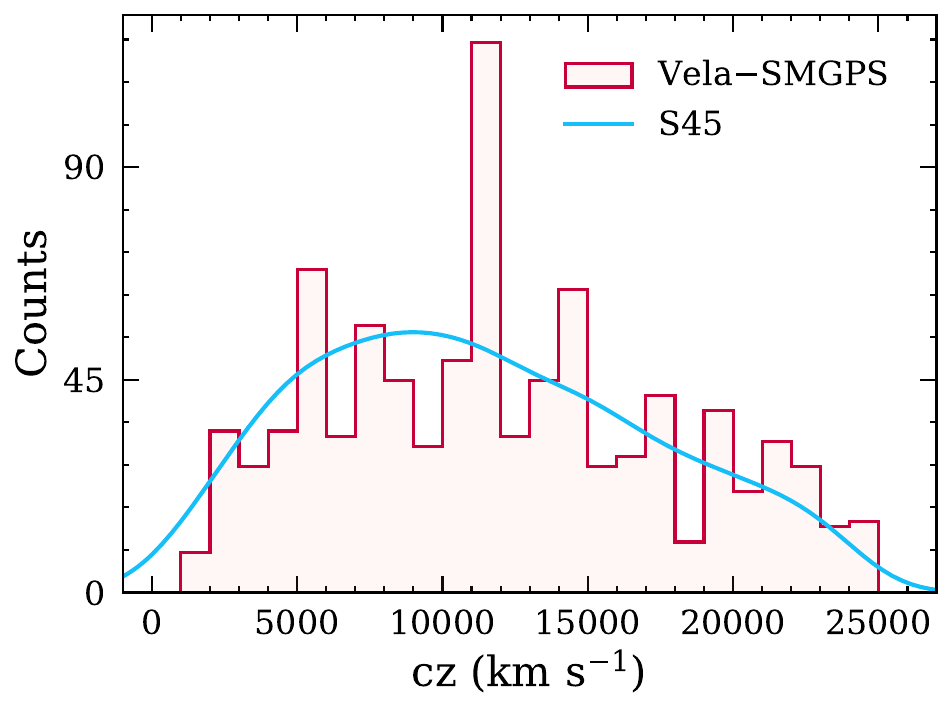}
	\centering
	\caption{ Velocity distribution for the full survey range (260$^{\circ} \leq \ell \leq 290^{\circ}$) of Vela$-$SMGPS detections (light-red histograms). The cyan lines correspond to the S45 simulations (rms $\sim 0.45$ \mJy{}).}
	\label{fig:vel_distribution}
	\end{figure}

\begin{figure*}
	\centering
	\includegraphics[width=\linewidth]{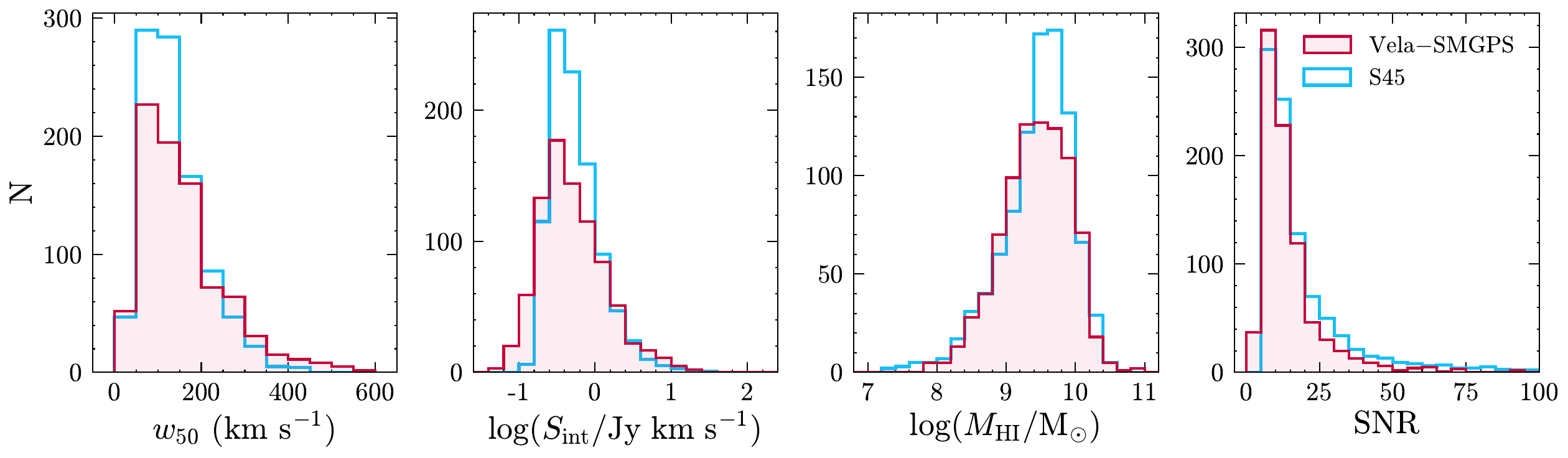}
	\centering
	\caption{ Distribution of global \hi{} parameters in Vela$-$SMGPS (light-red filled histogram) and S45 simulation (cyan histogram). The four panels display, from left to right, the linewidth ($w_{50}$) measured at the 50\% level of peak flux density, the logarithm of integrated flux ($S_{\mathrm{int}}$), the logarithm of \hi{} mass ($M_{\mathrm{HI}}$), and the integrated signal-to-noise ratio (SNR).}
	\label{fig:dist}
	\end{figure*}

We display the distribution of \hi{} properties for our detections with light-red filled histograms in Figs. \ref{fig:vel_distribution} and \ref{fig:dist}. The light blue lines represent the S45 simulations. 
We appropriately scaled the S45 counts per bin to match the surveyed area. The properties under comparison include velocities (Fig. \ref{fig:vel_distribution}), linewidths, integrated fluxes, \hi{} masses, and SNR (Fig. \ref{fig:dist}).  \\

\noindent
\emph{Velocity distribution.} Fig. \ref{fig:vel_distribution} visually presents an initial view of the overall velocity distribution in the survey area of Vela$-$SMGPS. The survey counts exhibit a shape quite similar to the simulations, with comparable median velocities: 11592 \kms{} (Vela$-$SMGPS) and 11209 \kms{} (S45). However, some peaks stand out, such as a prominent narrow peak around ${\sim}$12000 \kms{} with roughly 1.85 times higher counts than predicted. This feature is clearly seen as a strong wall-like feature in Fig. \ref{fig:wedge_hizoa}. Additionally, there is a hint of broader overdensity at the high-velocity end, i.e., $V_{\mathrm{hel}} > 18500 - 23000$ \kms{}.\\

\noindent 
\emph{$w_{50}$ linewidth distribution.} The linewidth histogram for Vela$-$SMGPS ranges from $12-615$ \kms{}. As expected from a systematic blind \hi{} survey (e.g., \citealt{Haynes2011,Staveley2016}), a strong prevalence at low-linewidth galaxies ($w_{50} \sim 50-200$ \kms{}) is observed. This trend is also confirmed by S45. Both distributions exhibit maximum peaks in the $50-100$ \kms{} bin range. We measure mean values of $158 \pm 99$ \kms{} and $137 \pm 71$ \kms{} for Vela$-$SMGPS and S45, respectively. The lower value for S45 reflects its higher proportion of low linewidth galaxies. On the other hand, the distribution in Vela$-$SMGPS shows more high-$w_{50}$ galaxies.
The deviations at the low and high linewidth distributions can be attributed to the inclusion of sources captured with their satellites, which were not spatially resolved in our data cubes. In contrast, S45 exclusively simulated unique galaxies.
\\

\noindent
\emph{Integrated} $H\scriptstyle{I}$ \emph{flux densities.} The second panel illustrates the distribution of the logarithm of the integrated flux. It ranges from a minimal value of $\log (S_{\rm int}/\rm Jy\,km\,s^{-1}) = -1.37$ ($S_{\rm int} = 0.04$ Jy \kms{}) to a maximum of $\log (S_{\rm int}/\rm Jy\,km\,s^{-1}) = 1.29$ ($S_{\rm int} = 19.43$ Jy \kms{}). Vela$-$SMGPS and S45 both peak at $\log (S_{\rm int}/\rm Jy\,km\,s^{-1}) \sim -0.5$. 
S45 presents a slightly higher mean value ($S_{\rm int} = 0.55$ Jy \kms{}) compared to Vela$-$SMGPS with a mean value of $S_{\rm int} = 0.5$ Jy \kms{}.
This slight difference is attributed to S45 excluding all galaxies with SNR\,$<5$. \\

\noindent
$H\scriptstyle{I}$ \emph{mass distribution}. In both Vela$-$SMGPS and S45, the galaxies span three orders of magnitude in \hi{} mass, ranging from $10^{7.8}\,$\msun\ to $10^{10.9}\,$\msun\ and $10^{7.1}\,$\msun\ to $10^{10.8}\,$\msun\, respectively. Apart from the more pronounced peaks at $\log (M_{\mathrm{HI}}/\rm M_{\odot}) \sim 9.5$ in Vela$-$SMGPS and 9.7 in S45, the overall distributions are very similar. Their mean \hi{} masses only differ by approximately 3\% with S45 being slightly higher. We find an average \hi{} mass of $10^{9.43}\,$\msun\ and $10^{9.44}\,$\msun\ for Vela$-$SMGPS and S45, respectively. These values are comparable to the measurement of the \hi{} mass of the Milky Way (see e.g., \citealt{Nakanishi2003}).\\

\noindent
\emph{Integrated signal-to-noise ratio.} In the rightmost panel, we compare the integrated signal-to-noise ratio between the simulation and the \hi{} detections from our survey, calculated using Eq. \ref{eq:snr}. Both distributions exhibit striking similarity, with a peak occurring at ${\sim}7\sigma$. As the SNR decreases, the number of detections increases steeply.\\

\begin{figure}
	\centering
	\includegraphics[width=\linewidth]{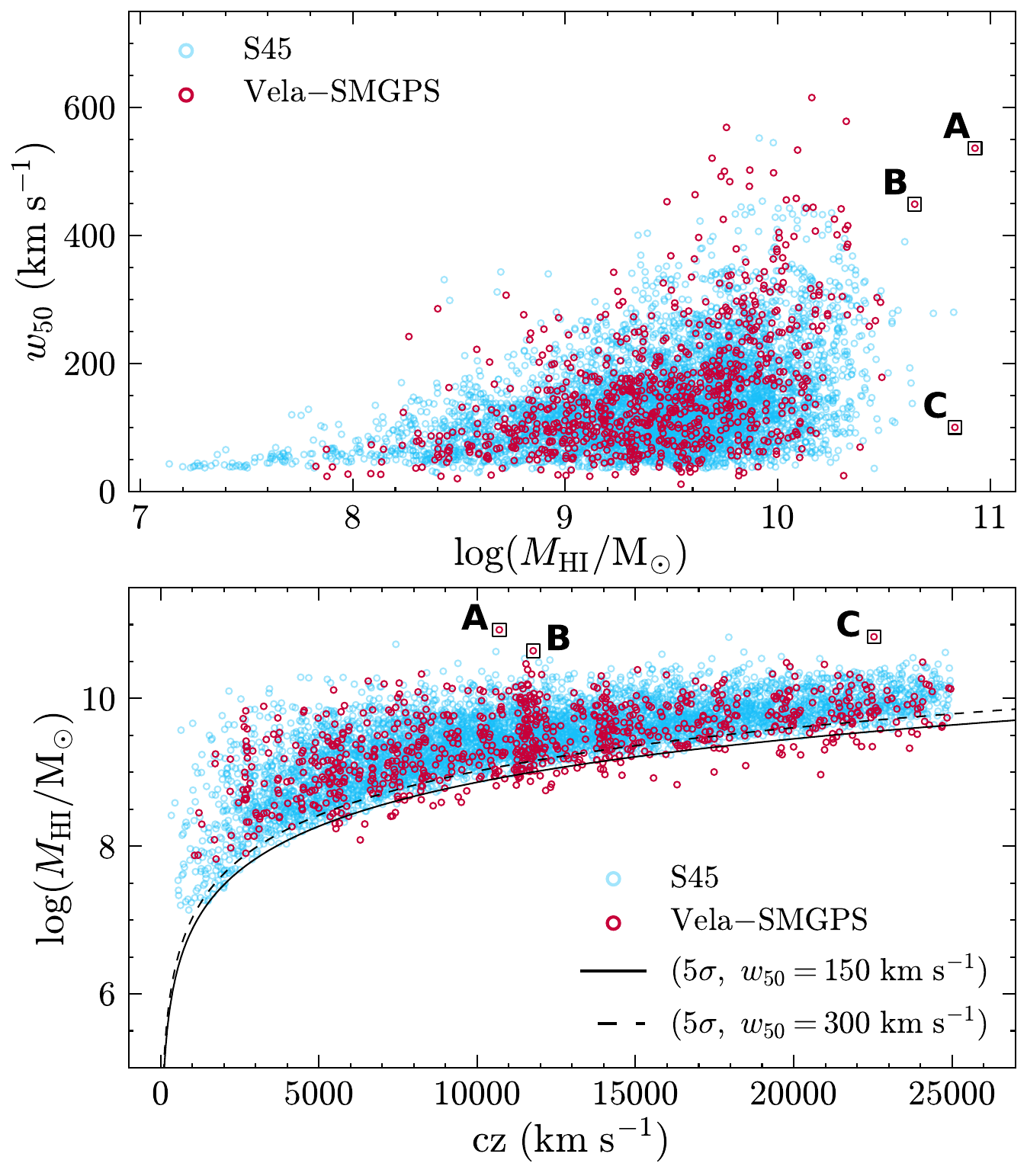}
	\centering
	\caption{ Dependence of \hi{} linewidth $w_{50}$ as a function of \hi{} mass (top panel) and \hi{} mass versus velocity (bottom panel). Vela$-$SMGPS detections (red dots) are overlaid on simulated S45 galaxies (cyan dots). The solid and dashed black lines represent sensitivity curves for the $5\sigma$ \hi{} mass detection level, assuming $w_{50}$ values of 150 and 300 \kms, with $\sigma = 0.45$ \mJy{}. SMGPS-HI J083650-433738 (A), J095513-561846 (B), and J084731-432134 (C) are marked inside black squares.}
        \vspace{-1.2em}
	\label{fig:sens_curve}
	\end{figure}

In Fig. \ref{fig:sens_curve} we compare the distribution of the individual detections in Vela$-$SMGPS (red dots) in $w_{50} - \log M_{\rm HI}$ and $\log M_{\rm HI} - V_{\rm hel}$ plots. We again add simulations to provide an idea about the expected distribution. The resulting distributions suggest that both the simulations and Vela$-$SMGPS sample the same population. Vela$-$SMGPS detects a slightly larger fraction of galaxies toward the high-mass end. 
We mark three galaxies seen as outliers in the figure due to their large \hi\ masses. The first one, SMGPS-HI J083650-433738 (marked as galaxy A), with $w_{50}$ = 536 \kms{} and $M_{\rm HI} = 10^{10.93}\,$\msun), is the most \hi{}-massive spiral recovered in HIPASS/HIZOA J0836-43, discussed earlier in Section \ref{sec:ancillary}. J095513-561846 (B) or HIZOA J0955-56, another broad linewidth galaxy with $w_{50}$ = 449 \kms{} and $M_{\rm HI} = 10^{10.65}\,$\msun. Another massive spiral galaxy, with a lower linewidth, J084731-432134 (C, $w_{50}$ = 100 \kms{}, $M_{\rm HI} = 10^{10.83}\,$\msun), was also detected as an outlier in Fig. \ref{fig:sens_curve} at $V_{\rm hel} \sim 22500$ \kms{}. It resembles a large face-on galaxy when examined in the Galaxy Atlas. 

The bottom panel of Fig. \ref{fig:sens_curve} shows the so-called sensitivity curve displaying $\log M_{\rm HI}$ vs. velocity. The black solid and dashed lines represent the 5$\sigma$ detection limit for galaxies with 150 \kms and 300 \kms\ linewidth as a function of redshift, respectively. First, Vela$-$SMGPS only found galaxies with masses $\log(M_{\rm HI}/\rm M_{\odot}) > 7.8$, whereas the simulation identifies galaxies below that threshold. This discrepancy may be attributed to the lower number of low linewidth galaxies (50 to 150 \kms) detected in Vela$-$SMGPS compared to S45 (see the left panel in Fig. \ref{fig:dist}). It could also be due to the coarse channel width or even LSS because we did not detect galaxies in the volume below $cz \leq 800$ \kms{}. Second, the distributions of detections and simulations both show that we are sensitive to normal spirals ($w_{50} \sim 200$ \kms{}) with $5\sigma$ $\log(M_{\rm HI}/\rm M_{\odot}) \geq 9.5$ at the VSCL distance of $V_{\rm hel} = 18000$ \kms{}, and $5\sigma ~ \log(M_{\rm HI}/\rm M_{\odot}) \geq 9.7$ out to the surveyed redshift limit of $V_{\rm hel} = 25000$ \kms{}. This demonstrates that we are sensitive to galaxies below $M^*_{\rm HI}$ at the VSCL distance.

The figure also highlights various signatures of LSS in its distribution. These high degrees of clustering seem to occur at various scales and are closely aligned to the hint of peak overdensities observed in the velocity distribution presented earlier in Fig \ref{fig:vel_distribution}. Four prominent clumps around 3000 \kms{}, 5000$-$6000 \kms{}, 11000$-$12000 \kms{}, and ${\sim}$14000 \kms{} are observed. Notably, the clump at 11000$-$12000 \kms{} exhibits an elongated distribution along the \hi{} mass. To examine this more closely, we will, in the next sections, divide the survey into different longitude ranges, and wedge diagrams, for a further analysis of the uncovered LSS. 

\section{Large Scale Structures Analysis} \label{sec:lss}

In this section, we will first inspect the velocity distribution of the survey detections in comparison to the simulation to assess peak overdensities, giving particular emphasis to areas where the VSCL walls are predicted to intersect. Subsequently, we examine the distribution of Vela$-$SMGPS galaxies in the inner ZOA to pinpoint these overdensities and assess their significance on on-sky plots and redshift slice diagrams.

\subsection{LSS in Velocity Space}\label{sec:overdens}

\begin{figure}
	\centering
	\includegraphics[width=0.9\linewidth]{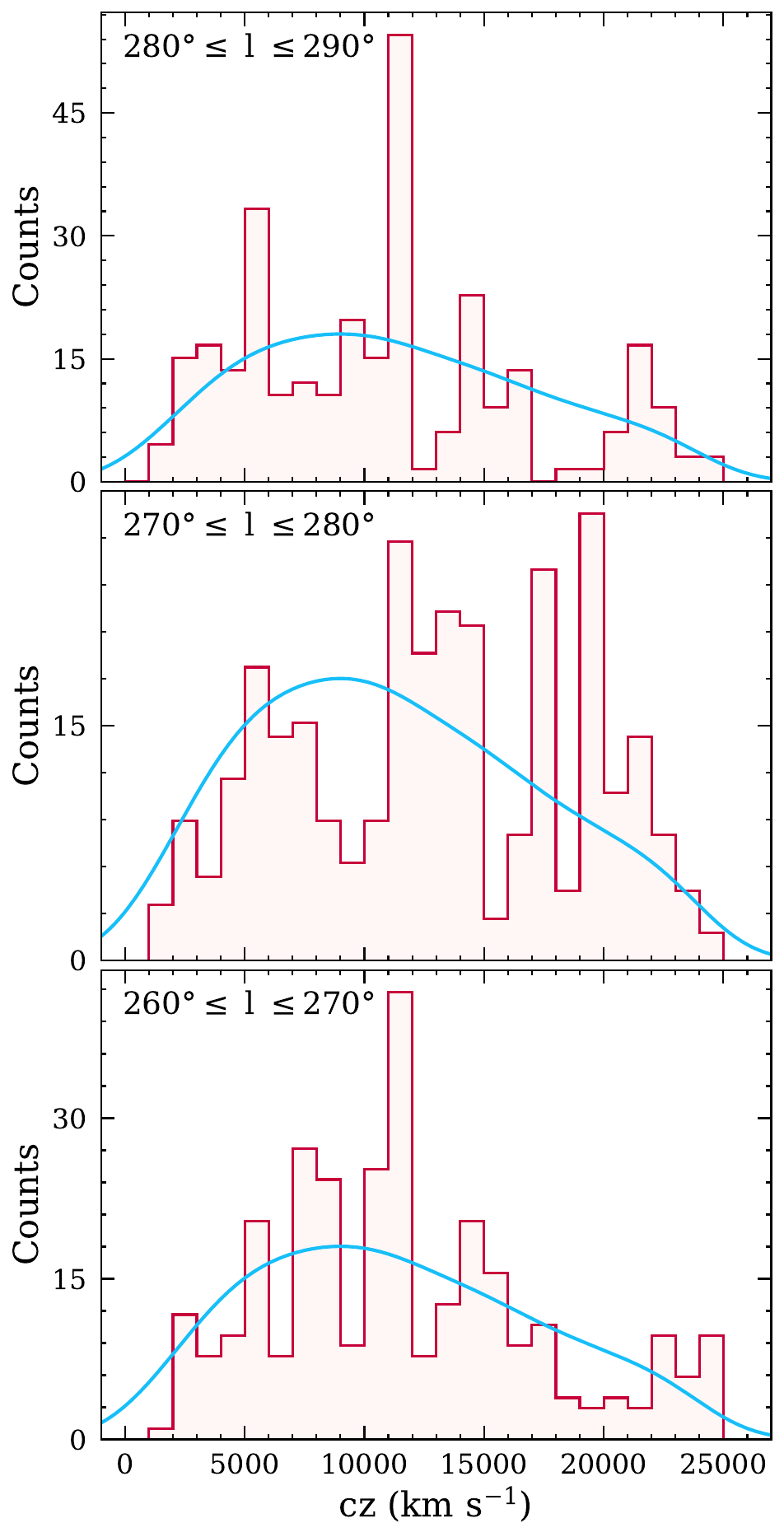}
	\centering
	\caption{ Velocity distribution histograms similar to Fig. \ref{fig:vel_distribution}, where the three different panels subsample distinct longitude ranges: 280$^{\circ} \leq \ell \leq 290^{\circ}$, 270$^{\circ} \leq \ell \leq 280^{\circ}$, and 260$^{\circ} \leq \ell \leq 270^{\circ}$.}
	\label{fig:overdensity}
	\end{figure}

In Section \ref{sec:properties}, we showed the overall velocity distribution of Vela$-$SMGPS (see Fig. \ref{fig:vel_distribution}) with potential peak overdensities. To localize these peaks and set a better understanding of these potential filaments and/or walls, we divide the survey area into different subregions with a longitude interval of $\Delta \ell = 10^{\circ}$ (see Fig. \ref{fig:overdensity}). 

The first panel of Fig. \ref{fig:overdensity} spans the range $280^{\circ} \leq \ell \leq 290^{\circ}$. We observe a subtle enhancement at $2000-4000$ \kms{} and a more pronounced peak at $5000-6000$ \kms{}, also notable in Fig. \ref{fig:sens_curve}. These hints are suggestive of links with the Hydra/Antlia wall \citep{kraan1995} and AS639 cluster \citep{Stein1996}, respectively (see Section \ref{sec:lss_structure} for further details). Most prominent is the significant peak at ${\sim}$12000 \kms{} which is also present in the other subregions. Its visibility is particularly pronounced in this first panel, with a factor of ${\sim}2$ higher than predicted counts. It is also quite conspicuous in the mass-redshift space (see bottom panel of Fig. \ref{fig:sens_curve}). The peak appears to be due to a narrow but very extended filament aligned along the survey's length in both longitude and latitude (see also Section \ref{sec:properties}).

The second panel covers the longitude range of $270^{\circ} \leq \ell \leq 280^{\circ}$. We see distinct features corresponding to higher redshift structures in Vela$-$SMGPS. This includes the significant clump observed at velocities of $14000-16000$ \kms{} in Fig. \ref{fig:sens_curve}, which is also visible in the other panels at velocities of $14000-15000$ \kms{}. Furthermore, two additional prominent peaks related to the VSCL $(V_{\mathrm{hel}} > 16000$ \kms{}) stand out over the predicted counts from S45. The counts are roughly 2.8 times higher than expected for the velocity range of $17000-18000$ \kms{}, and nearly 3.6 times higher for the range of $19000-20000$ \kms{}. This closely matches the longitude range ($270^{\circ}-280^{\circ}$) of the first Wall (W1) in VSCL, found from optical spectroscopy at latitudes $|b| \gtrsim  7^{\circ}$ (\citetalias{Kraan-Korteweg2017}). There is a hint of the second higher redshift Wall (W2) visible around $21000-22000$ \kms{}. The second Wall is, however, quite prominent in the last panel, $260^{\circ} \leq \ell \leq 270^{\circ}$, with counts that are roughly 2.5 times higher than those in the simulations.

\subsection{LSS crossing the ZOA}\label{sec:lss_structure}
We now look at the on-sky distributions of the newly discovered galaxies in a sequence of velocity intervals and discuss the unveiled LSS in connection to known LSS above and below the ZOA.  
Figures \ref{fig:LSS_ann} and \ref{fig:LSS2_ann} present a total of five on-sky (2D) distributions, each spanning a velocity range of $\Delta v = 4500$ \kms. The recently identified \hi{} galaxies are located within black dashed boxes which demarcate the survey area of Vela$-$SMGPS. Symbolized by cyan, green, and blue, each 1500 \kms{} bin represents the closest to the highest velocity bin within the respective panel. Grey contour lines, representing DIRBE/IRAS extinction contours \citep{Schlegel1998,Schlafly2011} at $A_{\mathrm{B}} = 3.\kern-0.5ex^{\mathrm{m}}0$, highlight regions where the ZOA is opaque to optical galaxies \citep[e.g.,][]{Woudt2001}. We display redshift wedges alongside the on-sky plots since galaxies that appear spatially aligned in the sky might not share the same alignment in redshift space. An investigation of both displays simultaneously allows a comprehensive examination of the galaxy clustering and density peaks seen in Figs. \ref{fig:sens_curve} and \ref{fig:overdensity}. \\

We will start from the nearest velocity bin in Fig. \ref{fig:LSS_ann}:

    (\emph{2000 $< V_{hel} < $ 6500 km s$^{-1}$}): a noticeable concentration of cyan dots ($2000-3500$ \kms{}) is centered around $\ell \sim 280^{\circ} \pm 4^{\circ}, b \sim 0^{\circ}$ (see red circle). 
    An alignment of blue points at ${\sim}3000$ \kms{} in the wedge diagram provides further support. This could be the signature associated with the filamentary structure known as the Hydra/Antlia extension \citep{kraan1995,Kraan2000_vela}, which crosses the ZOA at these coordinates.
    
    The upper right panel of the on-sky plot (\emph{6500 $< V_{hel} < $ 11000 km s$^{-1}$}) exhibits fewer pronounced features. Nevertheless, we observe a filament from $260^{\circ} \leq \ell \leq 277^{\circ}$ in the wedge diagram, which appears as a peak overdensity at $7000-9000$ \kms{} in the last panel of Fig. \ref{fig:overdensity}. 
    This structure may be linked to the cluster AS639 ($280^{\circ}, +6^{\circ}, 6000$ \kms{}; \citealt{Stein1996}).
    
    In the bottom left panel (\emph{11000 $< V_{hel} < $ 15500 km s$^{-1}$}), the distribution of galaxies in the velocity range of $11000-12500$ \kms{} (cyan dots) is most remarkable. It spans the entire longitude range of $\Delta \ell = 30^{\circ}$ within the survey area as seen in a highly significant peak present in all the panels of Fig. \ref{fig:overdensity}. The on-sky plot and wedge unveil a narrow filament at constant velocity, estimated to have a width of ${\sim}500$ \kms{} (${\sim}7$ Mpc) and an approximate length of $90$ Mpc at ${\sim}12000$ \kms{}. 
    Upon closer examination of the wedge slice, it looks like this structure could potentially consist of two intersecting filaments, possibly crossing between $270^{\circ} \leq \ell \leq 280^{\circ}$. 
    This observation is strengthened when looking at the on-sky plot displayed in the bottom left panel of Fig. \ref{fig:LSS_ann}, which shows Vela$-$SMGPS in context with known LSS data from the literature in the surrounding area ($245^{\circ}<\ell<305^{\circ}$, $|b|<25^{\circ}$). These include data from HyperLEDA \citep{Paturel2003}, 2MRS \citep{Huchra2012,Macri2019}, combined with unpublished redshift data from AAOmega+2dF, SALT, OPTOPUS, and 6dF in the Hydra/Antlia region (Kraan-Korteweg, priv. comm).
    One of the filaments is suggested to cross the GP around $275^{\circ} \leq \ell \leq 280^{\circ}$ (indicated by the clustering of green dots inside the red circle), continuing toward positive latitudes above the GP and slightly higher redshifts (blue dots).\\
    
We now turn our attention to the structures at the highest velocity intervals (\emph{15500 $< V_{hel} < $ 24500 km s$^{-1}$}), illustrated in Fig. \ref{fig:LSS2_ann}. This specific redshift incorporates the volume of the Vela Supercluster ($\ell = 272.5^{\circ} \pm 20^{\circ}, b = 0^{\circ} \pm 10^{\circ}, 19500 \pm 3500$ \kms{}, \citetalias{Kraan-Korteweg2017}). 

In the upper left panel (\emph{15500 $< V_{hel} < $ 20000 km s$^{-1}$}), a distinct clustering of green dots ($17000-18500$ \kms{}) is observed, concentrated primarily around $\ell \sim 270^{\circ}-277^{\circ}$ (light green rectangle). This marks the identified main VSCL Wall (W1). 
    Additionally, a segment of the second Wall (W2) at higher velocities (blue dots at $18500-20000$ \kms{}) is noted in a similar region ($\ell \sim 272^{\circ}-280^{\circ}$, indicated by the blue rectangle), with a higher density peak towards $\ell \sim 277^{\circ}$. 
    These clusterings are seen as density peaks when compared to simulations in the second panel of Fig. \ref{fig:overdensity}. 

    As we transition to the lower left panel (\emph{20000 $< V_{hel} < $ 24500 km s$^{-1}$}) to investigate W2, around $22000$ \kms, a consistent pattern emerges. The clustering of points, although more diffuse, indicates the presence of a broader and less compact wall. The cyan and green dots converge around $\ell \sim 272^{\circ}-278^{\circ}$, and $V_{\rm hel} \sim 20000-23000$ \kms{} (marked by the intersection of the cyan and green rectangles). This coincides with the smaller peaks in the last two panels of Fig. \ref{fig:overdensity}, less notable compared to W1.
    Interestingly, the representation of the walls in the wedge plot (see red arrows) strikingly mirrors the structures identified on both sides of the ZOA as seen in Fig. 3 of the VSCL discovery paper (\citetalias{Kraan-Korteweg2017}). \\

    As a result, despite only capturing a portion of VSCL due to the limited \hi{} mass sensitivity of $\log(M_{\rm HI}/\rm M_{\odot}) \sim 9.5$ at the VSCL distance, our findings are in support of the existence of clear overdensities associated with the surmised walls of the VSCL, W1 at ${\sim}18500$ \kms{} and W2 at ${\sim}21500$ \kms{}, and point in the approximate direction of the core of a substantial overdensity as derived from independent reconstructions \citep{Sorce2017,Courtois2019}. These structures seem to cross in the inner ZOA around $272^{\circ} \leq \ell \leq 278^{\circ}$ (see orange circle in the wedge), thus reinforcing the findings presented by \citetalias{Kraan-Korteweg2017}. 

    Even with the addition of new SMGPS data, the paucity of data at intermediate latitudes $(1.5^{\circ} \leq |b| \leq 7^{\circ})$ remains prominent in the on-sky plots, which makes it difficult to visualize the morphology of the walls and any potential merging process in detail. The forthcoming MeerKAT Vela$-$\hi{} survey (Rajohnson et al., in prep) will present newly obtained \hi{} data similar to the SMGPS. It will bridge the gap between the SMGPS and the optical spectroscopic data $(1.5^{\circ} \leq |b| \leq 7^{\circ})$ in the VSCL region. These will allow an expanded discussion into how these peaks of overdensity in the inner ZOA from Vela$-$SMGPS connect with known LSS at much higher latitudes above $|b| \gtrsim 7^{\circ}$.\\

\afterpage{%
\renewcommand\thefigure{\arabic{figure}a}
\onecolumn
\begin{landscape}
\begin{figure*}
    \centering
    \subfigure{\includegraphics[width=0.64\textwidth]{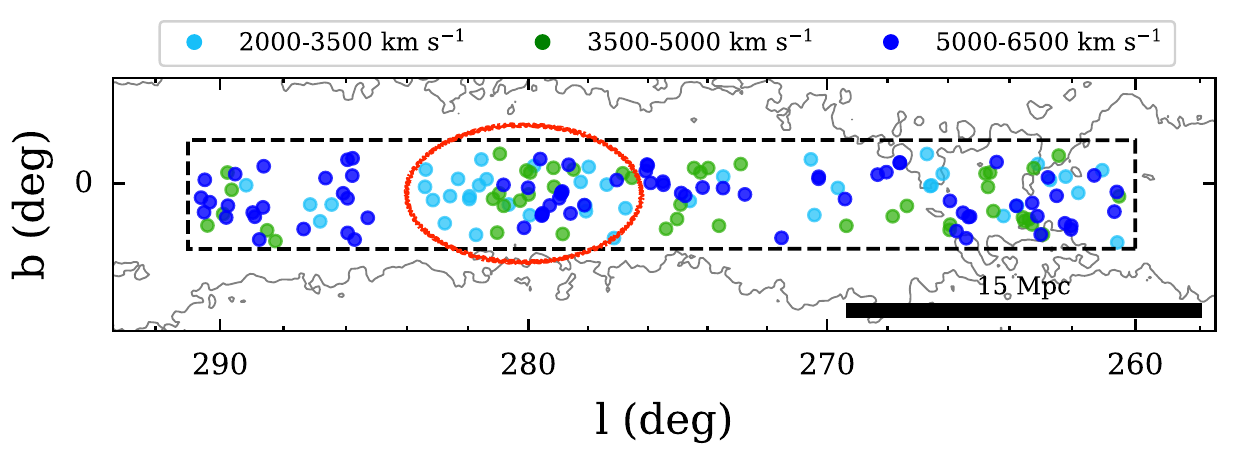}}%\hspace{2em}
    \subfigure{\includegraphics[width=0.61\textwidth]{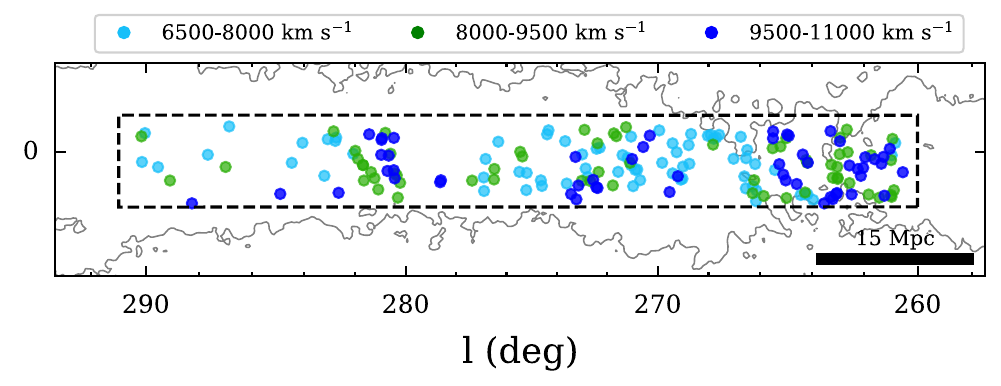}}\vspace{-2em}
    \subfigure{\includegraphics[width=0.64\textwidth]{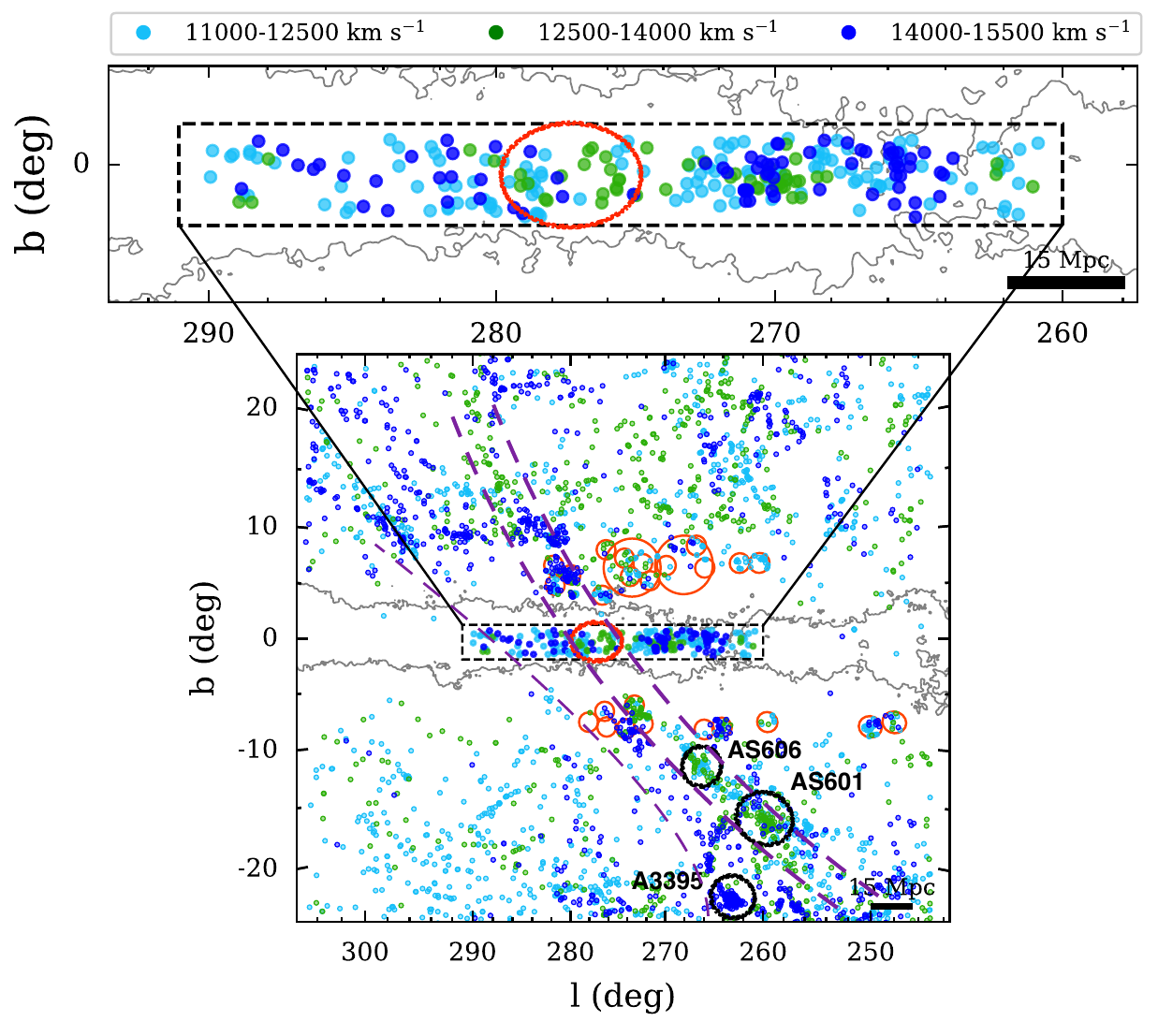}}\hspace{-0.5em}
    \subfigure{\includegraphics[width=0.63\textwidth]{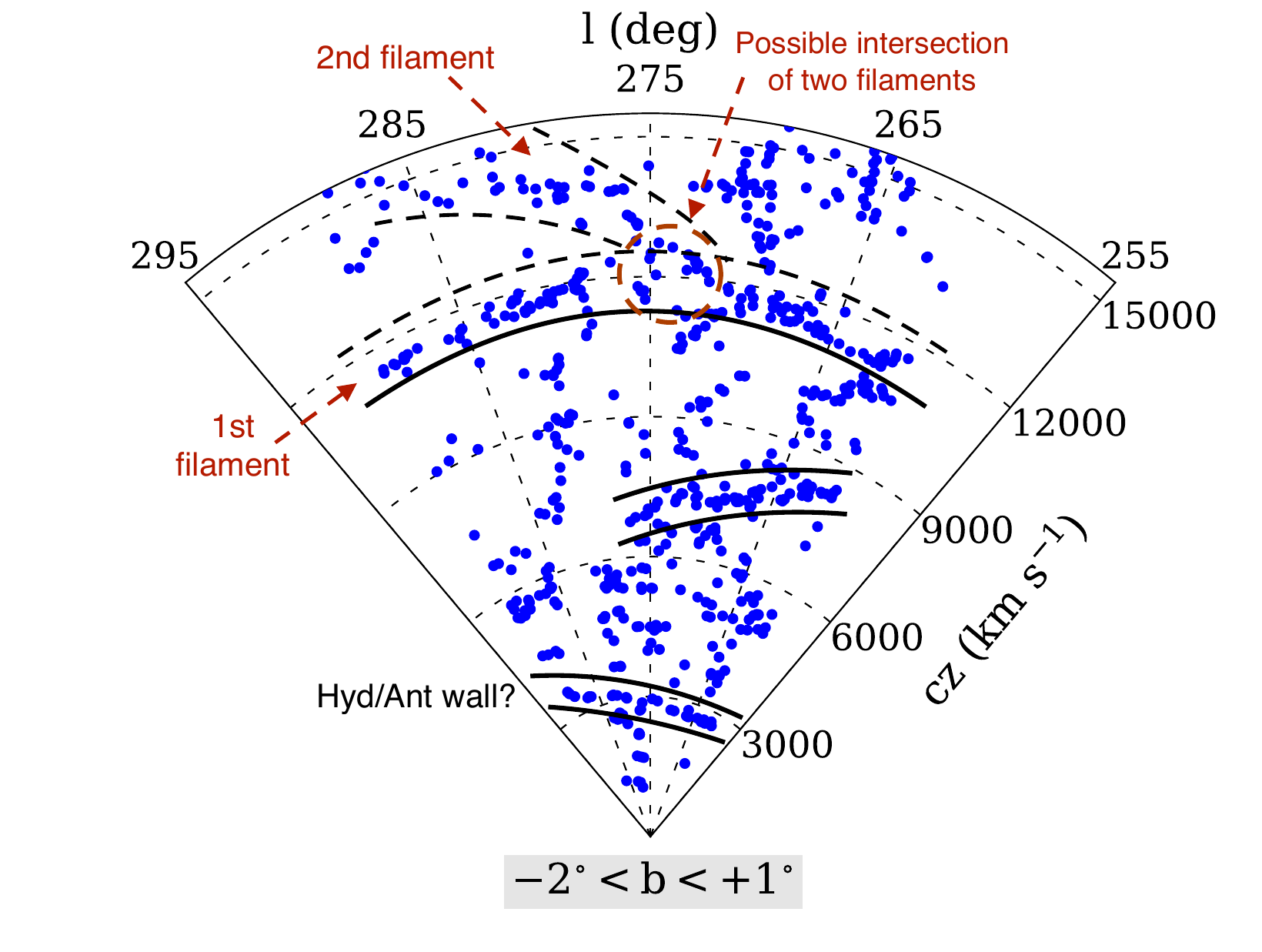}}
	\caption{ On-sky distributions of galaxies in the Vela$-$SMGPS survey, divided into three redshift intervals within the nearby velocity range $2000 < V_{\mathrm{hel}} < 15500$ \kms{}. Each panel spans $\Delta v = 4500$ \kms, with cyan, green, and blue indicating velocity bins increasing in steps of 1500 \kms{}. The black dashed box highlights the newly detected \hi{} galaxies. The bottom left panel (velocity interval $11000-15500$ \kms{}) is accompanied by the on-sky plot of its surrounding area ($245^{\circ}<\ell<305^{\circ}$, $|b|<25^{\circ}$), and includes ancillary data with known redshifts from HyperLEDA, 2MRS, and \citetalias{Kraan-Korteweg2017}. The latter is shown inside the orange open circles. Grey contour outlines dust extinction based on DIRBE maps \citep{Schlegel1998,Schlafly2011} at $A_{\mathrm{B}} = 3.\kern-0.5ex^{\mathrm{m}}0$. Each on-sky plot has a 15 Mpc scale bar at its bottom right corner. An accompanying wedge diagram limited to $V_{\mathrm{hel}} < 15500$ \kms{} is displayed in the bottom right panel.}
	\label{fig:LSS_ann}
\end{figure*}
\end{landscape}
\twocolumn
}

\afterpage{%
\renewcommand\thefigure{\arabic{figure}b}
\onecolumn
\begin{landscape}
\begin{figure*}
\ContinuedFloat
    \centering
    \subfigure{\includegraphics[width=0.65\textwidth]{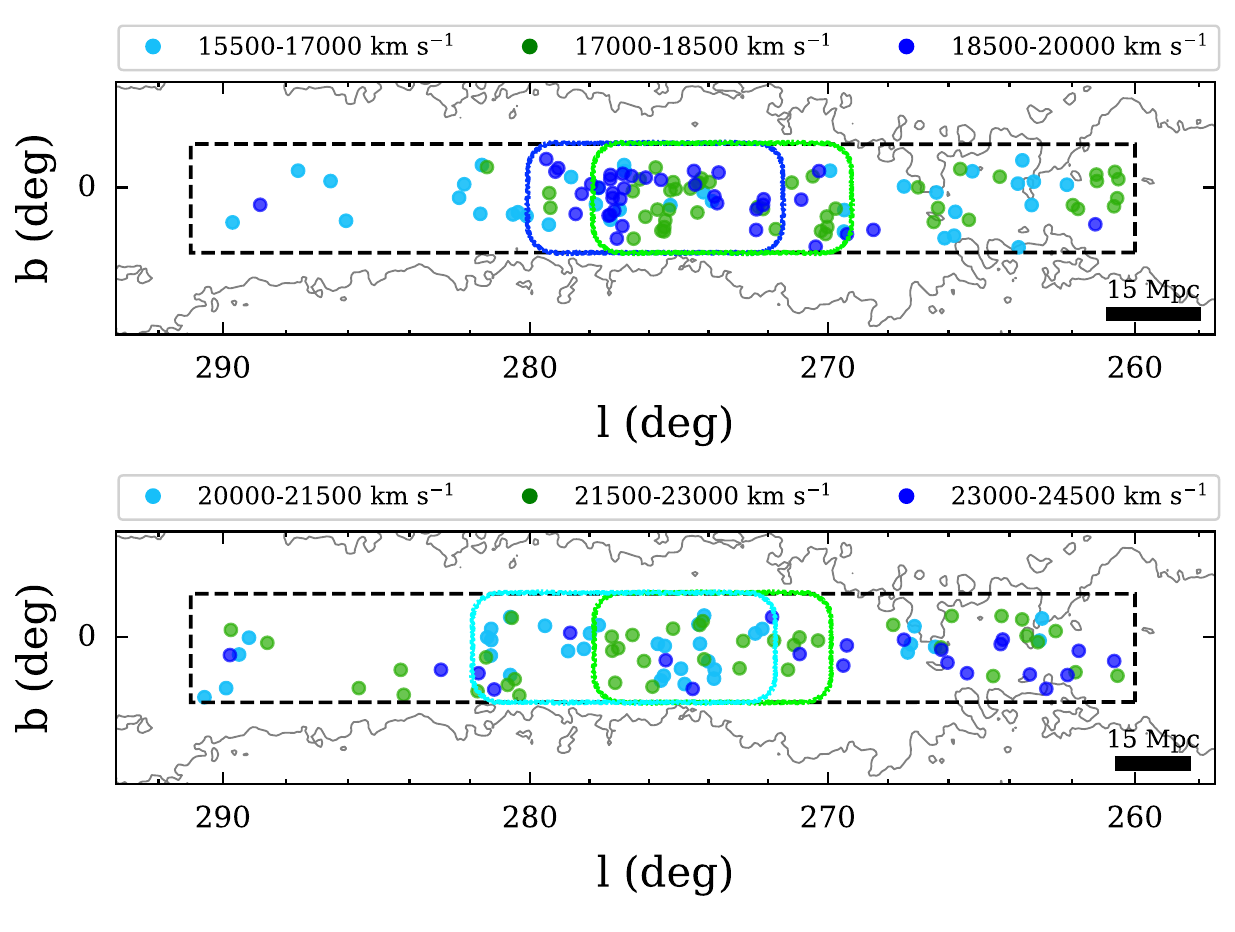}}
    \subfigure{\includegraphics[width=0.6\textwidth]{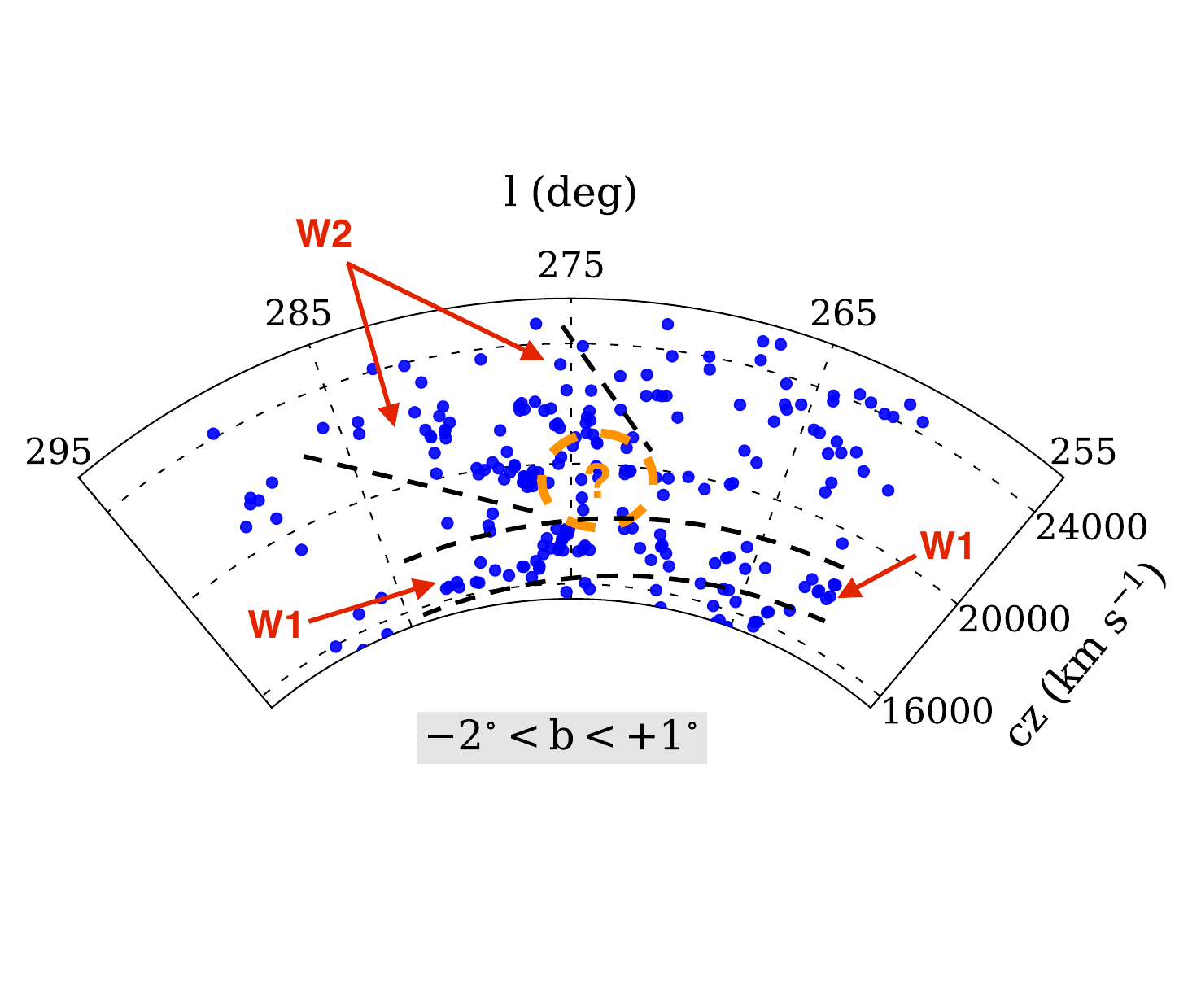}}
	\caption{ Continuing from Fig. \ref{fig:LSS_ann}, this figure presents two velocity bins with $\Delta v = 4500$ \kms in the left panels, covering the higher redshift range ($V_{\mathrm{hel}} > 15500$ \kms{}). The right panel features a wedge diagram within $15500 < V_{\mathrm{hel}} < 24500$ \kms{}.}
	\label{fig:LSS2_ann}
\end{figure*}
\end{landscape}
\twocolumn 
}

\section{Conclusions}\label{sec:conclusion}
This paper presents the \hi{} data extracted from the SARAO MeerKAT Galactic Plane Survey, centered on the Vela region (Vela$-$SMGPS; $260^{\circ} \leq \ell \leq 290^{\circ}, -2^{\circ} \leq b \leq 1^{\circ}$). With a velocity coverage reaching out to $V_\mathrm{hel} \sim 25000$ \kms{}, and a sensitivity of ${\sim}0.39$ \mJy{}, we have successfully cataloged galaxies along a narrow $3^{\circ}$ strip covering the most opaque part of the ZOA. Using SoFiA in conjunction with a visual verification process, we identified 843 \hi{} galaxies across the 10 contiguous mosaics covering the survey area. Among them, only 39 possessed ancillary redshift information for independent comparison, having previously been detected by the HIZOA survey. Six of these sources were found to consist of two or more counterparts (illustrated in Fig. \ref{fig:hiZOA_groups}). Notably, Vela$-$SMGPS successfully recovered the most \hi{}-massive spiral, HIZOA J0836-43, exhibiting consistent measurements with ATCA and HIZOA. The \hi{} properties of the galaxies in common are also in excellent agreement and exhibit a linear relation, consistent -- within the errors -- with a one-to-one relation over the entire range of the parameters. Following our expectations, we detect \hi{} sources well below the characteristic $M^*_{\rm HI}$ out to the edge of Vela$-$SMGPS.

Large-scale structures have been revealed from a comparison with expectations of a uniformly filled volume at a sensitivity threshold of 0.45 \mJy{} (S45). Our analysis pinpointed a high degree of clustering. Galaxy counts with a factor $\sim 2-4$ times higher compared to S45 were seen in the velocity ranges of $5000-6000$ \kms{} ($\ell \sim 280^{\circ}-290^{\circ}$), $11000-12000$ \kms{} (full $\Delta \ell \sim 30^{\circ}$), $17000-20000$ \kms{} and $21000-22000$ \kms{} ($\ell \sim 270^{\circ}-280^{\circ}$), and $22000-25000$ \kms{} ($\ell \sim 260^{\circ}-270^{\circ}$). The last two peaks of overdensity align with the position and redshift space of the previously identified two walls of the VSCL, providing strong evidence for the existence of their continuity within the surveyed volume. 
A thorough 3D distribution analysis of the notable density enhancements observed in the velocity distribution has resulted in several key observations of both various new structures, as well as confirmation of suspected links to known structures beyond the ZOA, such as:
\begin{itemize}
    \item[--] A distinct signature at 3500 \kms, substantiating the surmised Hydra/Antlia wall crossing into the ZOA at approximately $\ell \sim 280^{\circ}$. 
    \item[--] The identification of one or more intersecting filaments connecting structures both above and below the GP within $270^{\circ} \leq \ell \leq 280^{\circ}, V_\mathrm{hel} \sim 12000$ \kms{}. 
    \item[--] A glimpse of two wall-like overdensities in VSCL, with both W1 ($16000-19000$ \kms{}) and W2 ($19000-23000$ \kms{}) potentially crossing the GP at ${\sim}20000$ \kms{} within the longitude range of $272^{\circ} \leq \ell \leq 278^{\circ}$.\\
\end{itemize}

We found a previously uncharted population of highly obscured galaxies, particularly at the lowest latitudes and within an unprecedented redshift range for blind systematic \hi{} surveys in the ZOA.
These results clearly demonstrate the power of MeerKAT in providing a new window using systematic \hi\ surveys to explore the innermost ZOA despite extreme obscuration as well as continuum source contamination. 

In view of the success of this endeavor, we have launched a similar survey using MeerKAT for a wider systematic \hi{} survey to bridge the gap between Vela$-$SMGPS and previous research extending to higher latitudes. This survey has meanwhile been concluded. Its analysis together with the results presented here is expected to enable us to connect structures above and below the GP, fully trace the extent and morphology of the hidden part of VSCL, and measure its effect on the cosmic flow fields. 

\section*{Acknowledgements}
%MeerKAT, IDIA, CARTA, Caracal acknowledgments
We would like to extend our gratitude to Gyula J{\'o}zsa for his support in running CARACal, and Sarah White for her contributions to the development of the mosaicking module in CARACal.

The MeerKAT telescope is operated by the South African Radio Astronomy Observatory, which is a facility of the National Research Foundation, an agency of the Department of Science and Innovation. We acknowledge the use of the \href{www.ilifu.ac.za}{Ilifu} cloud computing facility, a partnership between the University of Cape Town, the University of the Western Cape, the University of Stellenbosch, Sol Plaatje University, the Cape Peninsula University of Technology and the South African Radio Astronomy Observatory. The Ilifu facility is supported by contributions from the Inter-University Institute for Data Intensive Astronomy (IDIA a partnership between the University of Cape Town, the University of Pretoria, the University of the Western Cape and the South African Radio Astronomy Observatory), the Computational Biology division at UCT and the Data Intensive Research Initiative of South Africa (DIRISA). We also applied the \href{https://cartavis.github.io}{CARTA} (Cube Analysis and Rendering Tool for Astronomy) software (DOI 10.5281/zenodo.3377984) in our analysis.

This paper makes use of the MeerKAT data with Project IDs: SSV-20180721-FC-01 and SCI-20180721-SP. The data published here have been reduced using the CARACal pipeline (\url{https://caracal.readthedocs.io/en/latest/}).

%Funding acknowledgments
SHAR, RCKK, HC, NS, SK, and DJP greatfully acknowledge support by the South African Research Chairs Initiative of the Department of Science and Technology and the National Research Foundation. 

%others:tools and databases.
We acknowledge the usage of the HyperLeda database (\url{http://leda.univ-lyon1.fr}). This research has made use of: the NASA/IPAC Extragalactic Database (NED), which is operated by the Jet Propulsion Laboratory, California Institute of Technology, under contract with the National Aeronautics and Space Administration; the NASA’s Astrophysics Data System Bibliographic Services; the VizieR catalogue access tool, CDS, Strasbourg, France (DOI:10.26093/cds/vizier). The original description of the VizieR service was published in 2000, A\&AS 143, 23. This publication utilizes data products from: the Two Micron All Sky Survey, which is a joint project of the University of Massachusetts and the Infrared Processing and Analysis Center/California Institute of Technology, funded by the National Aeronautics and Space Administration and the National Science Foundation; the Wide-field Infrared Survey Explorer, which is a joint project of the University of California, Los Angeles, and the Jet Propulsion Laboratory/California Institute of Technology, funded by the National Aeronautics and Space Administration. We also used \href{http://www.astropy.org}{Astropy}, a community-developed core Python package for Astronomy \citep{astropy:2013, astropy:2018}.

%%%%%%%%%%%%%%%%%%%%%%%%%%%%%%%%%%%%%%%%%%%%%%%%%%
\section*{Data Availability}

The complete Vela$-$SMGPS galaxy catalog is accessible as online ASCII supplementary material. Additionally, the Galaxy Atlas, featuring all Vela$-$SMGPS galaxy candidates, is available on the Zenodo repository: \url{https://doi.org/10.5281/zenodo.11160742}. The raw MeerKAT data are archived and can be accessed at \url{https://archive.sarao.ac.za}. Reduced mosaics can be obtained upon request.



\bibliographystyle{mnras}
\bibliography{reference} % if your bibtex file is called example.bib

%\begin{thebibliography}{99}
%\end{thebibliography}
%%%%%%%%%%%%%%%%%%%%%%%%%%%%%%%%%%%%%%%%%%%%%%%%%%

%%%%%%%%%%%%%%%%% APPENDICES %%%%%%%%%%%%%%%%%%%%%
\onecolumn
\appendix
\newpage
\twocolumn
\section{Sofia-2 parameter file}\label{app-A}

The appendix below provides the parameter file employed in SoFiA 2.3.1, using the smooth $+$ clip (S $+$ C) finder with a $4\sigma$ noise threshold. The same parameter configuration is utilized for the second SoFiA run but with the adjustment \texttt{scfind.threshold = 3.5}.

\begin{scriptsize}
\begin{verbatim}

# Global settings

pipeline.verbose           =  false
pipeline.pedantic          =  false

# Input

input.data                 =  Taper15_T30/to_galactic/T30_gal_taper15.fits
input.region               =  1268, 5470, 355, 3958, 9, 525 
input.gain                 =  
input.noise                =  Taper15_T30/to_galactic/T30_gal_taper15_noise.fits 
input.weights              =  
input.mask                 =  
input.invert               =  false

# Flagging

flag.region                =  
flag.auto                  =  false
flag.threshold             =  5.0
flag.log                   =  true

# Noise scaling

scaleNoise.enable          =  true
scaleNoise.mode            = local 
scaleNoise.statistic       =  mad
scaleNoise.windowXY        =  301 
scaleNoise.windowZ         =  15 
scaleNoise.gridXY          =  0
scaleNoise.gridZ           =  0
scaleNoise.interpolate     =  true 
scaleNoise.scfind          =  false 

# S+C finder (smooth & clip)

scfind.enable              =  true
scfind.kernelsXY           =  0, 7, 15
scfind.kernelsZ            =  0, 3, 5, 7
scfind.threshold           =  4 #and 3.5
scfind.replacement         =  1.5 
scfind.statistic           =  mad
scfind.fluxRange           =  negative

# Linker

linker.radiusXY            =  5
linker.radiusZ             =  3 
linker.minSizeXY           =  10
linker.minSizeZ            =  3
linker.maxSizeXY           =  0
linker.maxSizeZ            =  50

# Reliability

reliability.enable         =  true
reliability.threshold      =  0.90 
reliability.scaleKernel    =  0.25
reliability.minSNR         =  3
reliability.plot           =  true
reliability.debug          =  true

# Parameterisation

parameter.enable           =  true
parameter.wcs              =  true
parameter.physical         =  true
parameter.prefix           =  SoFiA
parameter.offset           =  true 

# Output

output.directory           =  Taper15_T30/sofia_new/local/4_sigma
output.filename            =  T30_gal_local_4_sigma_
output.writeCatASCII       =  true
output.writeCatXML         =  false
output.writeCatSQL         =  false
output.writeNoise          =  false
output.writeFiltered       =  false
output.writeMask           =  true
output.writeMask2d         =  false
output.writeMoments        =  false 
output.writeCubelets       =  false 
output.marginCubelets      =  0
output.overwrite           =  true
\end{verbatim}
\end{scriptsize}

\pagebreak
\onecolumn 
\section{HIZOA detections in Vela$-$SMGPS}\label{app-B}
The 39 Vela$-$SMGPS detections in common with HIZOA \citep{Staveley2016} are shown in Fig. \ref{fig:hiZOA_unique1} (unique) and \ref{fig:hiZOA_groups} (multiple counterparts).

\begin{wrapfigure}{c}{1\textwidth}
	\centering
        \vspace{1cm}
	\includegraphics[width=\linewidth]{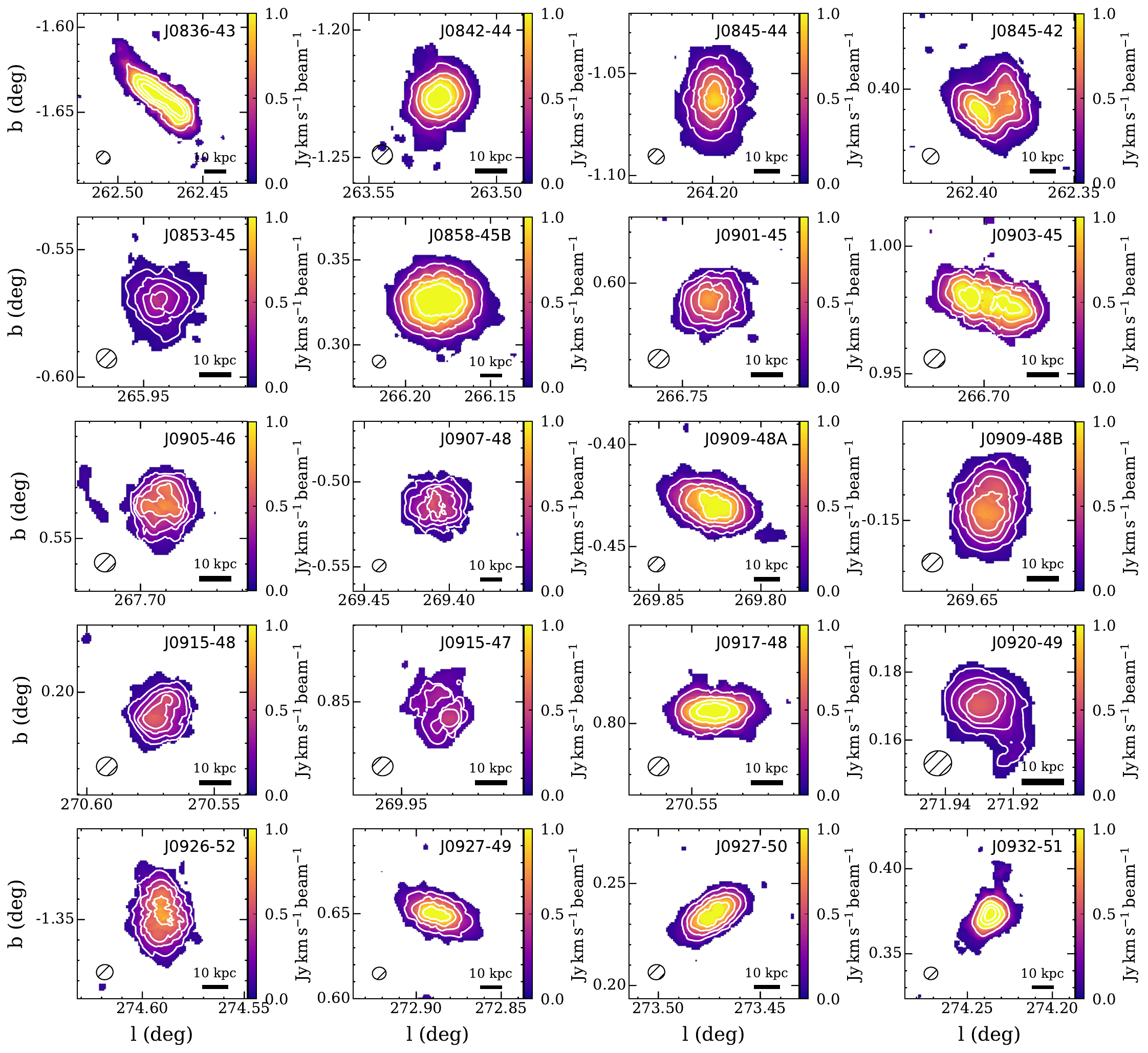}
	\centering
        \captionsetup{singlelinecheck=off,justification=centering,font=small,labelfont=bf,labelsep=period}
	\caption{ Moment-0 maps of HIZOA galaxies detected in Vela$-$SMGPS. The HIZOA identification name, the beam size, and a 10 kpc scale are shown in the upper right, bottom left, and bottom right corners of each image, respectively.}
	\label{fig:hiZOA_unique1}
\end{wrapfigure}

\textcolor{white}{White space test} 

\newpage
\FloatBarrier	
\begin{figure*}
    \ContinuedFloat
    \centering
    \includegraphics[width=\linewidth]{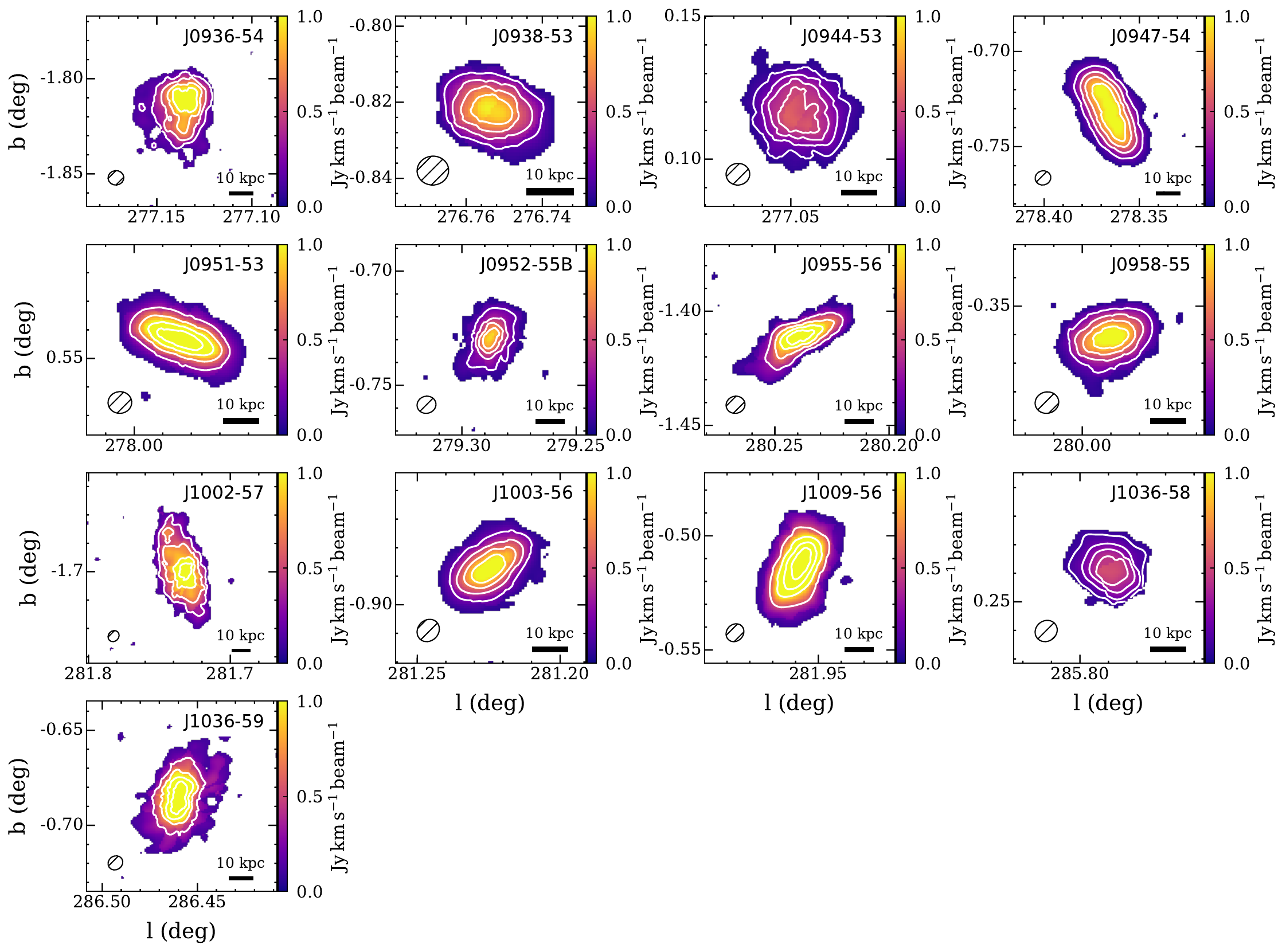}
	\caption{ Moment-0 maps of HIZOA galaxies detected in Vela-SMGPS (Continued).}
	\label{fig:hiZOA_unique2}
\end{figure*}

\begin{figure*}
	\centering
	\includegraphics[width=0.9\linewidth]{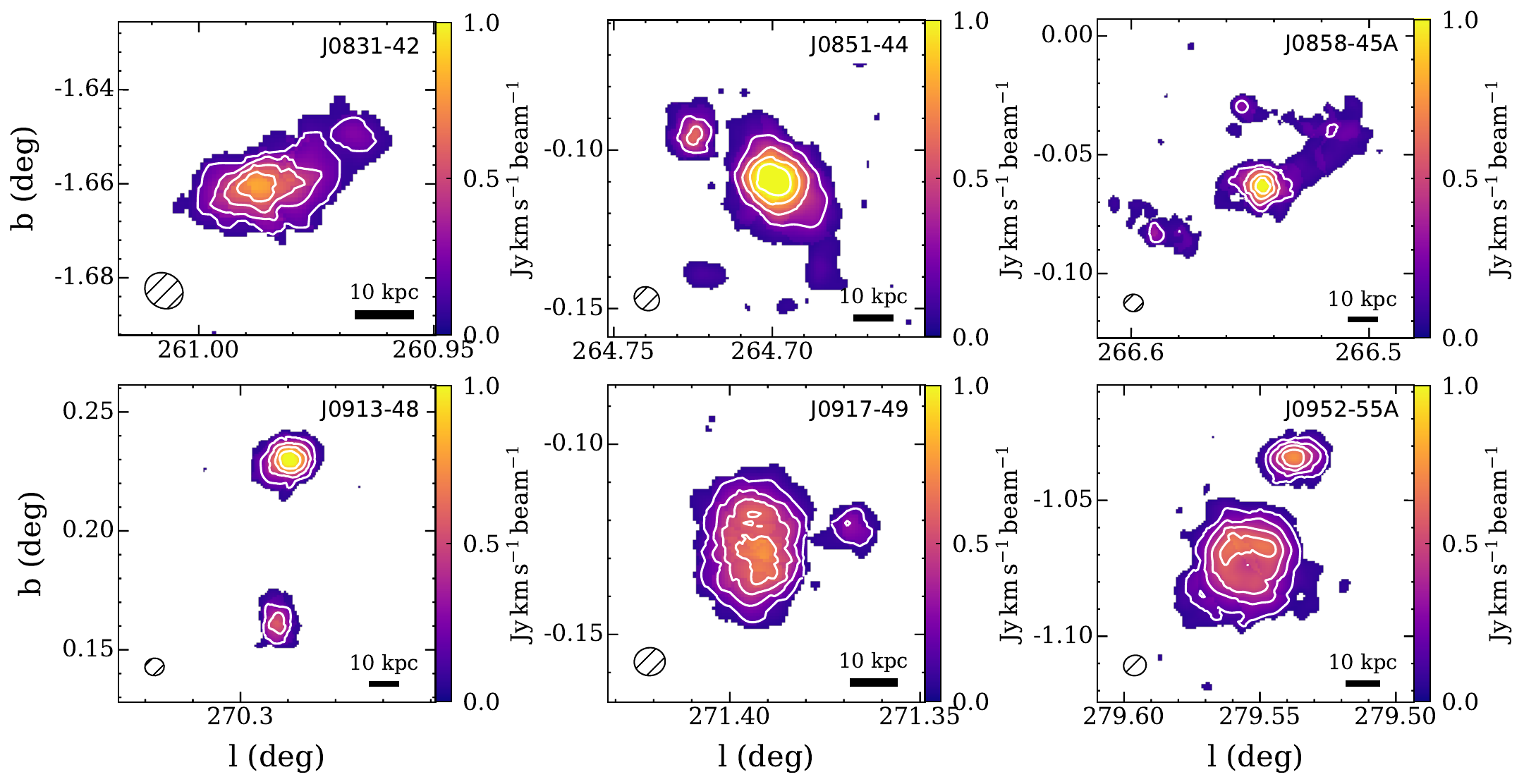}
	\centering
	\caption{ Moment-0 maps of HIZOA galaxies resolved into several galaxies in Vela$-$SMGPS.}
	\label{fig:hiZOA_groups}
	\end{figure*}
\FloatBarrier

\pagebreak
\section{Vela$-$SMGPS Full catalog}\label{app-D}
In Table \ref{tab:cat}, we provide the derived \hi{} properties obtained for the detected Vela$-$SMGPS detections ($260^\circ \leq \ell \leq 290^\circ$). A brief description of the columns in the catalog is provided below:\\

\textit{Column 1:} Galaxy identifier (ID): Prefixed with SMGPS-HI followed by J\textsc{hhmmss-ddmmss} representing the rounded J2000 equatorial coordinates (RA and DEC) of the source. 

\textit{Column 2:} The Vela$-$SMGPS Mosaic (T\textsc{xx}) in which the source has been detected (see e.g top panel of Fig. \ref{fig:GPS_full}) [T24$-$T33].

\textit{Columns 3 and 4:} Galactic coordinates $\ell$ and $b$ in units of degrees.

\textit{Column 5:} Peak flux density $S_{\rm peak}$ as derived from the Global \hi{} profile in Jy.

\textit{Columns 6 and 7:} Integrated flux $S_{\rm int}$ with its corresponding error ($\varepsilon_s$) in Jy \kms 
(see Eq. \ref{eq:int_flux} in Section \ref{subsec:parameterization}).

\textit{Column 8:} Mean local rms in \mJy{}. 

\textit{Column 9:} Optical heliocentric systemic velocity $V_\mathrm{hel}$ in \kms.

\textit{Columns 10 and 11:} Linewidths ($w_{20}$, $w_{50}$) in \kms. These represent the measured linewidths at which the profile reaches 20\% and 50\% of its peak flux density, corrected for instrumental broadening (see Eq. \ref{eq:width} and Eq. \ref{eq:w50_corr}).

\textit{Column 12:} Logarithm of the \hi\ mass determined from Eq. \ref{eq:hi_mass} [$\log (M_{\rm HI}$/\msun)]. 

\textit{Column 13:} Flags 1 or 2, where 1 signifies that the detection was classified as solid, 2 for a possible detection according to the visual verification process of each source (see Section \ref{subsec:classification}). An asterisk ($^*$) next to the flag code means that the detection has also been found in the area with the adjacent mosaic.  

\textit{Column 14:} Notes describing whether a galaxy is interacting (\emph{i}), presents one or multiple companions (\emph{w.comp(s)}), exhibits a wiggly baseline in its profile (\emph{BL}) or if the measurement of $w_{20}$ is uncertain due to the baseline behaviour (\emph{$w_{20}$ unc.}). Additionally, galaxies detected in major \hi{} and IR catalogs are marked.  For galaxies found in HIZOA \citep{Staveley2016}, their object names (e.g., J0831-42) are provided. `H' denotes \hi{} counterparts for galaxies also detected in HIPASS \citep{Meyer2004}. `I' refers to the IRAS Point Source Catalog \citep{Helou1988}, `M' stands for 2MASX \citep{Jarrett2000}, `W' represents matches from the Wide-field Infrared Survey Explorer catalog (WISE, \citealt{Cutri2014}), and `Z' signifies Deep NIR follow-up observations of HIZOA galaxies obtained with the IRSF Telescope \citep{Williams2014,Said2016}.

\clearpage
\onecolumn %reset to one column
%\begin{table*}
\begingroup
{\scriptsize
\centering
%\begin{threeparttable}
{\setlength\tabcolsep{4pt}
\begin{ThreePartTable}
\begin{TableNotes}
  \item[*] Detected by the overlapping region of the next mosaic as well.
\end{TableNotes}
\begin{longtable}{ccrrccccrrrrrl}
    \captionsetup{singlelinecheck=off,justification=centering,labelfont=bf,labelsep=period}
    \caption{\textbf{Vela-SMGPS full galaxy catalog: \hi{} properties.}}
	\label{tab:cat}\\
    \hline
    \hline
           \multicolumn{1}{c}{Name} & \multicolumn{1}{c}{Mosaic} &        \multicolumn{1}{c}{$\ell$} &      \multicolumn{1}{c}{$b$} & 
           \multicolumn{1}{c}{$S_{\rm peak}$} &
           \multicolumn{1}{c}{$S_{\rm int}$} &  \multicolumn{1}{c}{$\varepsilon_s$} &  \multicolumn{1}{c}{rms} &  $V_\mathrm{hel}$ &  $w_{20}$ &  $w_{50}$ &  $M_{\mathrm{HI}}$ & \multicolumn{1}{c}{Flag} &                               \multicolumn{1}{c}{Note} \\
           
            J2000 & TXX  &        \multicolumn{1}{c}{deg} &      \multicolumn{1}{c}{deg} & \multicolumn{1}{c}{Jy} &  Jy \kms{} &  \multicolumn{1}{c}{Jy \kms{}} &  \multicolumn{1}{c}{\mJy{}} &  \kms &  \kms &  \kms &  $\log (M_{\rm HI}$/\msun) & 1/2  &   \\
           
           \multicolumn{1}{c}{(1)} & \multicolumn{1}{c}{(2)} &        \multicolumn{1}{c}{(3)} &      \multicolumn{1}{c}{(4)} &  (5) &  \multicolumn{1}{c}{(6)} &  \multicolumn{1}{c}{(7)} &  (8) &  (9) &  (10) &  (11) & \multicolumn{1}{c}{(12)} &                               \multicolumn{1}{c}{(13)} & \multicolumn{1}{c}{(14)} \\
    \midrule
    \endfirsthead
    \caption*{Table \ref{tab:cat} continued}\\
    \hline
    \hline
    \multicolumn{1}{c}{Name} & \multicolumn{1}{c}{Mosaic} &        \multicolumn{1}{c}{$l$} &      \multicolumn{1}{c}{$b$} & 
           \multicolumn{1}{c}{$S_{\rm peak}$} &
           \multicolumn{1}{c}{$S_{\rm int}$} &  \multicolumn{1}{c}{$\varepsilon_s$} &  \multicolumn{1}{c}{rms} &  $V_\mathrm{hel}$ &  $w_{20}$ &  $w_{50}$ &  $M_{\mathrm{HI}}$ & \multicolumn{1}{c}{Flag} &                               \multicolumn{1}{c}{Note} \\
           
            J2000 & TXX  &        \multicolumn{1}{c}{deg} &      \multicolumn{1}{c}{deg} & \multicolumn{1}{c}{Jy} &  Jy \kms{} &  \multicolumn{1}{c}{Jy \kms{}} &  \multicolumn{1}{c}{\mJy{}} &  \kms &  \kms &  \kms &  $\log (M_{\rm HI}$/\msun) & 1/2  &   \\
           
           \multicolumn{1}{c}{(1)} & \multicolumn{1}{c}{(2)} &        \multicolumn{1}{c}{(3)} &      \multicolumn{1}{c}{(4)} &  (5) &  \multicolumn{1}{c}{(6)} &  \multicolumn{1}{c}{(7)} &  (8) &  (9) &  (10) &  (11) & \multicolumn{1}{c}{(12)} &                               \multicolumn{1}{c}{(13)} & \multicolumn{1}{c}{(14)} \\
    \midrule
    \endhead % all the lines above this will be repeated on every page
    \hline
    \insertTableNotes  % tell LaTeX where to insert the contents of "TableNotes"
    \endfoot %hline at the bottom of each table
    \hline
    \endlastfoot

J082930-421453 &    T33 &  260.566 & -1.903 &  0.010 &   1.25 &      0.23 &       0.46 &   2710 &  209 &   86 &      8.65 &    1 &                                               \\
 J083133-422815 &    T33 &  260.970 & -1.726 &  0.018 &   6.45 &      0.64 &       0.40 &   8242 &  466 &  415 &     10.33 &    1 &    J0831-42? w.comp (J083154-422631), M, W, Z \\
 J083154-422631 &    T33 &  260.983 & -1.659 &  0.018 &   2.99 &      0.42 &       0.36 &   8347 &  235 &  171 &     10.01 &    1 &          J0831-42, i, w.comp (J083133-422815) \\
 J083217-415102 &    T33 &  260.548 & -1.251 &  0.005 &   0.63 &      0.06 &       0.23 &  21824 &  315 &  277 &     10.18 &    1 &                                               \\
 J083223-421453 &    T33 &  260.881 & -1.471 &  0.003 &   0.26 &      0.05 &       0.27 &   8427 &  251 &  200 &      8.95 &    2 &                                               \\
 J083236-423939 &    T33 &  261.237 & -1.683 &  0.008 &   1.26 &      0.18 &       0.40 &  10767 &  448 &  330 &      9.85 &    1 &                                               \\
 J083253-425142 &    T33 &  261.429 & -1.762 &  0.006 &   1.22 &      0.17 &       0.43 &   8251 &  396 &  363 &      9.61 &    1 &                                          M, W \\
 J083312-425407 &    T33 &  261.497 & -1.737 &  0.011 &   2.28 &      0.22 &       0.39 &  11612 &  337 &  284 &     10.18 &    1 &                                               \\
\bottomrule
\end{longtable}
%\end{tabular}
\end{ThreePartTable}
}}

\clearpage
\twocolumn
 %for arxiv
%%%%%%%%%%%%%%%%%%%%%%%%%%%%%%%%%%%%%%%%%%%%%%%%%%
% Don't change these lines
\bsp	% typesetting comment
\label{lastpage}
\end{document}